\documentclass[twocolumn,english,aps,amssymb,amsmath,pra,showpacs,superscriptaddress,final]{revtex4}
\usepackage{ae,aecompl}
\usepackage[T1]{fontenc}
\usepackage[latin9]{inputenc}
\usepackage{amsmath}
\usepackage{graphicx}
\usepackage{amssymb}

\makeatletter
\@ifundefined{textcolor}{}
{%
 \definecolor{BLACK}{gray}{0}
 \definecolor{WHITE}{gray}{1}
 \definecolor{RED}{rgb}{1,0,0}
 \definecolor{GREEN}{rgb}{0,1,0}
 \definecolor{BLUE}{rgb}{0,0,1}
 \definecolor{CYAN}{cmyk}{1,0,0,0}
 \definecolor{MAGENTA}{cmyk}{0,1,0,0}
 \definecolor{YELLOW}{cmyk}{0,0,1,0}
 }

\@ifundefined{definecolor}
 {\usepackage{color}}{}
\@ifundefined{definecolor}
 {\@ifundefined{definecolor}
 {\usepackage{color}}{}
}{}

\usepackage{dcolumn}
\usepackage{bm}
\usepackage{subfigure}

\usepackage{amsfonts}\usepackage{latexsym}
\usepackage[matrix,frame,arrow]{xy}
\usepackage{epsf}

\makeatother

\usepackage{babel}

\makeatother

\makeatother

\makeatother

\usepackage{babel}

\begin{document}

\title{Optimal pulse spacing for dynamical decoupling in the presence of
a purely-dephasing spin-bath}

\author{Ashok Ajoy}

\email{ashok.ajoy@gmail.com}

\affiliation{Fakultät Physik, Technische Universität Dortmund, D-44221 Dortmund,
Germany.}

\affiliation{Birla Institute of Technology and Science - Pilani, Zuarinagar, Goa
- 403726, India.}

\affiliation{NMR Research Centre, Indian Institute of Science, Bangalore - 560012,
India.}

\author{Gonzalo A. Álvarez}

\email{galvarez@e3.physik.uni-dortmund.de}

\affiliation{Fakultät Physik, Technische Universität Dortmund, D-44221 Dortmund,
Germany.}

\author{Dieter Suter}

\email{Dieter.Suter@tu-dortmund.de}

\affiliation{Fakultät Physik, Technische Universität Dortmund, D-44221 Dortmund,
Germany.}

\keywords{decoherence, spin dynamics, NMR, quantum computation, quantum information
processing, dynamical decoupling, quantum memories }

\pacs{03.65.Yz,03.67.Pp,76.60.-k ,76.60.Lz }
\begin{abstract}
Maintaining quantum coherence is a crucial requirement for quantum
computation; hence protecting quantum systems against their irreversible
corruption due to environmental noise is an important open problem.
Dynamical decoupling (DD) is an effective method for reducing decoherence
with a low control overhead. It also plays an important role in quantum
metrology, where for instance it is employed in multiparameter estimation.
While a sequence of equidistant control pulses (CPMG) has been ubiquitously
used for decoupling, Uhrig recently proposed that a non-equidistant
pulse sequence (UDD) may enhance DD performance, especially for systems
where the spectral density of the environment has a sharp frequency
cutoff. On the other hand, equidistant sequences outperform UDD for
soft cutoffs. The relative advantage provided by UDD for intermediate
regimes is not clear. In this paper, we analyze the relative DD performance
in this regime experimentally, using solid-state nuclear magnetic
resonance. Our system-qubits are $^{13}$C nuclear spins and the environment
consists of a $^{1}$H nuclear spin-bath whose spectral density is
close to a normal (Gaussian) distribution. We find that in the presence
of such a bath, the CPMG sequence outperforms the UDD sequence. An
analogy between dynamical decoupling and interference effects in optics
provides an intuitive explanation as to why the CPMG sequence performs
superior to any non-equidistant DD sequence in the presence of this
kind of environmental noise. 
\end{abstract}
\maketitle

\section{Introduction}


Quantum information processing (QIP) relies on storing and manipulating
information in quantum mechanical states associated with accessible
subunits called qubits \cite{Nielsen00}. For reliable QIP, the information
encoded in the quantum register (the collection of qubits) must be
retained for arbitrarily long time \cite{Preskill1998}. However,
in any physically realizable QIP architecture, the qubits cannot be
completely isolated, but are weakly coupled to a large number of degrees
of freedom of their environment (bath). This causes the corruption
of the quantum state associated with the qubit, a process known as
decoherence \cite{Zurek03}. This process limits the time scale over
which quantum information can be retained \cite{Zurek03} and the
distance over which it can be transmitted \cite{Chiara2005,Burrell2007,Allcock2009,AlvarezPRL2010}.

Combating this decoherence process to extend the life time of quantum
states or processes and the distance bounds to transmit them is a
necessary step in building a quantum computer \cite{Shor1994}. One
of the simplest and most effective techniques suggested for this purpose
is Dynamical Decoupling (DD) \cite{5916,Yang2010}. It also has several
applications in quantum metrology, for example in the magnetometry
using single spins \cite{Taylor2008,Hall2010,Lange2010magn} and multiparameter
estimation \cite{Boixo2008}. DD consists of the application of $\pi$-pulses
to the qubits, which revert the decay due to the system-environment
interaction. The simplest implementation of this method is the Hahn
echo experiment \cite{Hahn1950}. It is employed in the case when
the system-environment (SE) interaction is a pure dephasing process
-- i.e. there is no net exchange of energy between the system and
the bath.

However, it is immediately obvious that any component of the interaction
with the bath that varies on a time scale shorter than $\tau$, the
delay between the pulses, cannot be refocused \cite{Carr1954,Meiboom1958}.
Hence the effectiveness of the Hahn echo and its generalization as
a train of equidistant pulses -- the CP \cite{Carr1954} and CPMG
\cite{Meiboom1958} sequences -- depend crucially on keeping the delays
$\tau$ between the pulses sufficiently short. However, this delay
is always limited by hardware or the maximum power deposition in the
sample. Moreover, rapid DD pulsing interferes with the controls that
are necessary for computation \cite{viola1999universal,viola2003}.
For example, if one wants to choreograph the dynamics of the spins
in a particular manner while still requiring that coherence survives
for long, the requisite controls have to be applied in parallel with
the DD pulses. Although the recent proposal of Dynamically Corrected
Gates (DCGs) \cite{Khodjasteh2009,Khodjasteh2010,west_high_2009,west_near-optimal_2010}
seeks to address this problem, when $\tau$ is very short there is
an enormous demand on the hardware.

Hence there is a strong motivation for finding a DD protocol that,
for a given number of pulses, provides the best performance. For the
case of a purely dephasing SE interaction, a possible approach to
this problem consists in visualizing the DD pulses as generating a
filter for the SE interaction (environmental modes) \cite{Cywinski2008,Biercuk2010}.
For example, the CPMG sequence acts as a band-stop or high-pass filter
(with bandwidth $\omega=2\pi/\tau$) since any interaction component
that varies slower than $\tau$ is refocused and filtered out. The
resulting decay rate of the quantum state is determined by the overlap
of the spectral density of the bath with the filter generated by the
pulse sequence \cite{Cywinski2008,Biercuk2010}. If one knows the
exact form of this function, it may be possible to design a suitable
filter, and hence a DD sequence, that leads to the slowest decay \cite{Uhrig2007,Gordon2008,Biercuk2009a,Uys2009,Clausen2010,pan2010,Biercuk2010}.
This was the motivation behind the DD sequence suggested by Uhrig
\cite{Uhrig2007}. It provides the non-equidistant distribution of
pulses that causes the flattest band-stop filter around $\omega=0$
\cite{Cywinski2008,Uhrig2008}, and is considered the best sequence
for combating low frequency noise. Experimental work \cite{Biercuk2009a,Biercuk2009,Jenista2009,Du2009}
showed that indeed UDD outperforms CPMG for certain types of noise
with a sharp high-frequency cut off.

However, when the spectral density of the bath has a long tail (soft
cutoff), the CPMG sequence has been predicted \cite{Cywinski2008,pasini_optimized_2010}
and found \cite{Biercuk2009a,Biercuk2009} to outperform the UDD sequence.
In an intermediate regime, for example a spin-bath where the spectral
density is Gaussian \cite{Abragam}, it was shown recently that CPMG
outperforms the UDD sequence \cite{Alvarez2010,Lange2010,Barthel2010,Ryan2010}.
However, a rigorous comparison of both these sequences in such regimes
is a matter of current research \cite{UhrigLidar2010,Hodgson2010,chen2010,viola2010,Biercuk2010}.
Of particular interest is determining the conditions (for example,
the maximum allowable $\tau$) under which the UDD continues to provide
an advantage over CPMG.

This was tackled in a purely dephasing bosonic bath by Hodgson \emph{et
al.} \cite{Hodgson2010}. They predicted that DD protocols with non-equidistant
pulses like UDD lose their advantages when lower bounds exist for
the pulse separation $\tau$. Uhrig and Lidar \cite{Uhrig2010} presented
analytical performance bounds for UDD. They considered instantaneous
perfect pulses and bounded environment and generic system-environment
Hamiltonians. They showed that for a fixed total duration, the survival
probability can be increased by increasing the number of pulses by
incrementing the DD order. However, if a minimum delay between pulses
is imposed, because the sequence time scales with the number of pulses
(DD order), they predict that longer cycles (higher orders) are not
always advantageous. These predictions are based on perturbative treatments
of the SE interaction. Accordingly, they apply to the fast control
regime where the cycle time is sufficiently short for the perturbative
treatment. Because these approaches cannot lead to a lower bound on
the attainable DD error in the presence of timing constraints, recently
Khodjasteh \emph{et al.} \cite{viola2010} obtained bounds for the
UDD performance as a function of its order by using an alternative
method for bosonic baths.

In light of this, it is important to provide a fair and practically
relevant comparison between these different DD approaches. Here, we
use the benchmarking methodology of our previous work \cite{Alvarez2010}:
which DD sequence maximizes the survival time of the quantum coherence
for a given average spacing between the pulses? In the earlier paper,
we concentrated on a comparison between different sequences with equidistant
pulses; here, we analyze the effect of a variable pulse spacing, using
the same approach. The spectral density of our bath is roughly Gaussian,
which may be considered to be intermediate between a sharp cutoff
and a long tail.

Several recent experimental results \cite{Alvarez2010,Lange2010,Barthel2010,Ryan2010}
indicate that CPMG performs better than UDD when the qubit is coupled
to a spin-bath. The superior CPMG performance has been attributed
to the fact that it compensates pulse errors along the direction of
the rotation axis \cite{Barthel2010,Ryan2010}, while such compensation
is reduced in UDD \cite{Biercuk2009a,Biercuk2009}. Also it was attributed
to a possible soft-cut off of their model system \cite{Lange2010,Barthel2010,Ryan2010}.
Our experiments and simulations indicate that even in the case of
ideal pulses, the CPMG outperforms UDD. In fact, we conjecture that
for a purely dephasing SE interaction, when the spectral density is
a Gaussian distribution and a rapidly fluctuating environment, the
CPMG sequence would maintain the state of the qubit system for the
longest period. Our reasoning is based on determining how the DD pulse
sequence alters the time-averaged SE interaction that the system {}``sees'',
for each frequency component of this interaction \cite{Cywinski2008}.
Drawing an analogy with optics, each pulse of the Uhrig sequence modifies
the SE interaction such that the interactions in different windows
{}``interfere'' destructively with each other. This destructive
interference is perfect for slow interactions. While the UDD protocol
provides the \textit{flattest} stop-band at $\omega=0$ \cite{Cywinski2008},
the CPMG protocol provides the \textit{widest}. When a DD sequence
is applied repetitively (say $M$ times), it becomes analogous to
passing the different frequency components of the SE interaction through
a diffraction grating. Overall, the optical analogy presented in this
paper provides an intuitive way to understand how the shape of the
spectral density function affects DD performance.

This paper is organized as follows. Sec. \ref{sec:The-system} describes
the physical system that we use in our experiments. In Sec. \ref{sec:dd}
we review the key ideas of dynamical decoupling. Sec. \ref{sec:pulses}
provides the pulse sequence protocols of UDD \cite{Uhrig2007,Uhrig2008}
and CPMG \cite{Carr1954,Meiboom1958}. Sec. \ref{sec:expts} contains
the results of our experiments. Sec. \ref{sec:physics} describes
a semi-classical model that explains the experimental results, and
suggests that the CPMG sequence is the best DD sequence for combating
a purely-dephasing SE interaction in a spin-bath. Within this section
we draw an optical analogy to DD sequences.

\section{The system\label{sec:The-system}}

\label{sec:system-2} For our experimental evaluation of the relative
performance of DD sequences, we use the simplest possible system consisting
of a single qubit and a purely dephasing environment consisting of
a spin-bath. For the system qubit, we use $^{13}$C nuclear spins
($S=1/2$), and for the environment $^{1}$H nuclear spins ($I=1/2$)
that act as the spin-bath. We use polycrystalline adamantane where
the natural abundance of the $^{13}$C nuclei is about 1\%, and to
a good approximation each $^{13}$C nuclear spin is surrounded by
about 100 $^{1}$H nuclear spins. The $^{13}$C-$^{13}$C interaction
is completely negligible compared to the $^{13}$C-$^{1}$H and $^{1}$H-$^{1}$H
interactions. The spin-spin interaction is dominated by the dipolar
interaction \cite{Abragam}.

We shall refer to the spin operator of the system qubit ($^{13}$C)
as $\hat{S}$, and to the $j^{th}$ bath spin ($^{1}$H) as $\hat{I}^{j}$.
The Zeeman frequencies of the system- and bath spins are $\omega_{S}$
and $\omega_{I}$, respectively. The dipolar coupling constants for
the system-bath interaction are $b_{Sj}$, and for the intra-bath
interaction $d_{ij}$.

It is convenient to describe the dynamics of the system in a rotating
frame of reference \cite{Abragam}, where the system rotates at the
(angular) frequency $\omega_{S}$ around the $z$-axis and the environment
at $\omega_{I}$. The total (free evolution) Hamiltonian is then \begin{equation}
\widehat{\mathcal{H}}_{f}=\widehat{\mathcal{H}}_{S}+\widehat{\mathcal{H}}_{SE}+\widehat{\mathcal{H}}_{E},\label{eq:Hfreeevolution}\end{equation}
 where $\widehat{\mathcal{H}}_{S}$ is the system Hamiltonian, $\widehat{\mathcal{H}}_{E}$
the environment Hamiltonian and $\widehat{\mathcal{H}}_{SE}$ the
system-environment interaction Hamiltonian: \begin{eqnarray}
\widehat{\mathcal{H}}_{S} & = & \hat{0},\label{eq:HS}\\
\widehat{\mathcal{H}}_{SE} & = & \hat{S}_{z}\sum_{j}b_{Sj}\hat{I}_{z}^{j},\label{eq:HSE}\\
\widehat{\mathcal{H}}_{E} & = & \sum_{i<j}d_{ij}\left[2\hat{I}_{z}^{i}\hat{I}_{z}^{j}-(\hat{I}_{x}^{i}\hat{I}_{x}^{j}+\hat{I}_{y}^{i}\hat{I}_{y}^{j})\right].\label{eq:HE}\end{eqnarray}

This type of system is encountered in a wide range of solid-state
spin systems, as for example electron spins in diamonds \cite{Naydenov2010,Lange2010,Ryan2010},
electron spins in quantum dots \cite{Bluhm2010,Barthel2010,Hanson2007}
and donors in silicon \cite{Kane1998,Morton2008}, which appear to
be promising candidates for future QIP implementations. In particular,
we consider the case where the interaction with the bath is weak or
comparable with the intra-bath interaction.

\section{Dynamical decoupling\label{sec:Dynamical-decoupling}}

\label{sec:dd} %
\begin{figure}
\includegraphics[bb=15bp 473bp 530bp 707bp,scale=0.4]{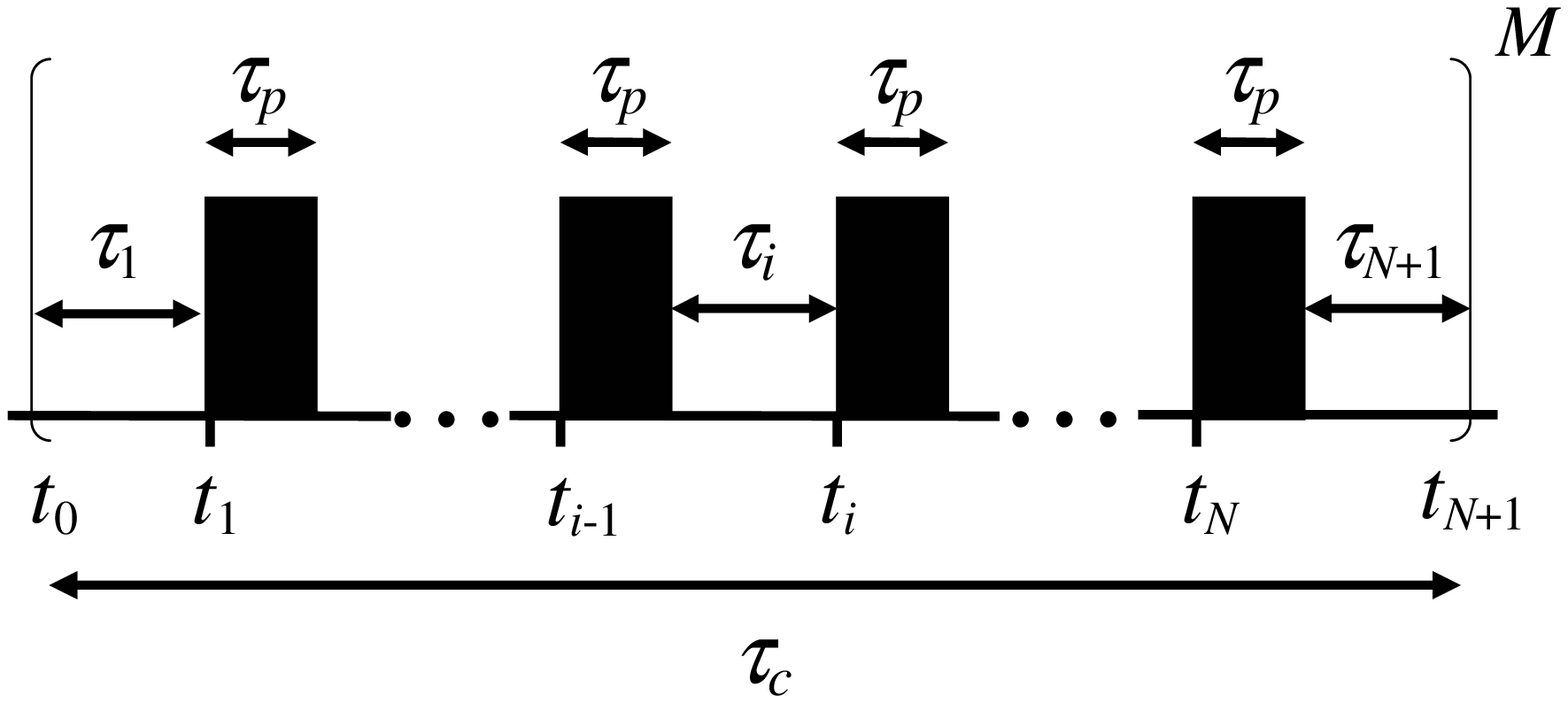}

\caption{Schematic representation of dynamical decoupling. The solid boxes
represents the control pulses.}

\label{Flo:DDScheme} %
\end{figure}

Dynamical decoupling aims to reduce the interaction of the system
with the environment by applying sequences of short, strong pulses
that invert the system qubits \cite{5916,Yang2010}. We write $\widehat{\mathcal{H}}_{{\rm \textrm{C}(S)}}(t)$
for the corresponding control Hamiltonians. It is assumed that the
environment cannot be directly controlled.

DD sequences usually consist of cycles of pulses. Figure \ref{Flo:DDScheme}
shows an example of such a cycle. In the rotating frame, the operator
that describes the evolution of the total system from 0 to $\tau_{c}$,
the duration of the cycle, is \begin{equation}
\hat{U}\left(\tau_{c}\right)=\hat{U}_{f}\left(\tau_{N+1}\right)\prod_{i=1}^{N}\hat{U}_{C}^{i}\left(\tau_{p}\right)\hat{U}_{f}\left(\tau_{i}\right),\label{eq:TotalU}\end{equation}
 where the free evolution operator is \begin{eqnarray}
\hat{U}_{f}\left(t\right) & = & \exp\left\{ -\mbox{i}\widehat{\mathcal{H}}_{f}t\right\} \end{eqnarray}
 and the control evolution operators are\begin{eqnarray}
\hat{U}_{C}^{i}\left(\tau_{p}\right) & = & T\exp\left\{ -i\int_{0}^{\tau_{p}}dt^{\prime}\left(\widehat{\mathcal{H}}_{f}+\widehat{\mathcal{H}}_{{\rm \textrm{C}(S)}}^{i}(t^{\prime})\right)\right\} ,\label{eq:UC}\end{eqnarray}
 where $T$ is the Dyson time-ordering operator \cite{dyson0,dyson1}
and $\tau_{p}$ the pulse duration. Let $t_{i}$ represent the time
at which the $i^{th}$ control operation starts. Then, the delay times
between the control Hamiltonians are $\tau_{i}=t_{i}-\left(t_{i-1}+\tau_{p}\right)$
for $i=2,..,n+1$ and $\tau_{1}=t_{1}-t_{0}$, where $t_{0}=0$ and
$t_{N+1}=\tau_{c}$. If the basic cycle is iterated $M$ times (see
Fig. \ref{Flo:DDScheme}), the total evolution operator becomes \begin{equation}
\hat{U}\left(t=M\tau_{c}\right)=\left[\hat{U}\left(\tau_{c}\right)\right]^{M}.\end{equation}

\begin{figure}
\includegraphics[scale=0.35]{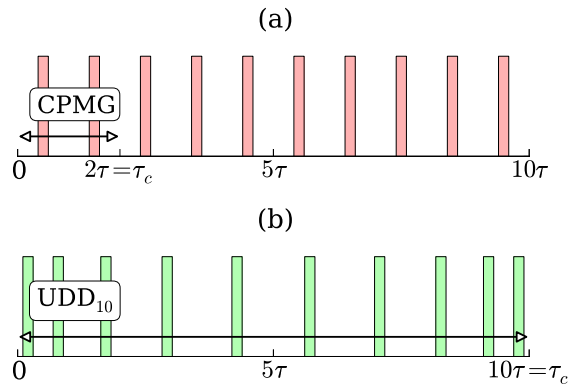}

\caption{(Color online) Distribution of pulses for (a) CPMG and (b) UDD$_{10}$
sequences. Here $\tau$ is the average distance between pulses. The
arrows denote the length of a single cycle of period $\tau_{c}$.
Note that a CPMG cycle consists of 2 pulses while an $N$$^{th}$order
UDD cycle has $N$ pulses.}

\label{Flo:Pulse_Seq} %
\end{figure}

The propagator (\ref{eq:TotalU}) for a single cycle of duration $\tau_{c}$
can be written in terms of an effective Hamiltonian: \begin{equation}
\hat{U}\left(\tau_{c}\right)=e^{-i\widehat{\mathcal{H}}_{eff}\tau_{c}}.\end{equation}
 Using average Hamiltonian theory \cite{Haeberlen1976} $\widehat{\mathcal{H}}_{eff}$
can be expressed as a series expansion, using, e.g., the Magnus expansion
\cite{Magnus1954}, \begin{align}
\widehat{\mathcal{H}}_{eff} & =\widehat{\mathcal{H}}^{(0)}+\widehat{\mathcal{H}}^{(1)}+\widehat{\mathcal{H}}^{(2)}+...=\sum_{m=0}^{\infty}\widehat{\mathcal{H}}^{(m)}.\end{align}
 The lowest order term $\widehat{\mathcal{H}}^{(0)}$ is the average
over the period $\tau_{c}$, \begin{equation}
\widehat{\mathcal{H}}^{(0)}=\frac{1}{\tau_{c}}\int_{0}^{\tau_{c}}\widehat{\mathcal{H}}(t)dt.\end{equation}
 A DD sequence with ideal pulses makes $\widehat{\mathcal{H}}^{(0)}=\widehat{\mathcal{H}}_{E}$
, i.e., the interaction Hamiltonian vanishes to zeroth order. In addition,
the DD sequences are designed such that the higher order terms have
reduced norm or vanish \cite{khodjasteh_fault-tolerant_2005,Uhrig2007,Yang2010}.

As we discussed in the previous section, the $^{13}$C-$^{13}$C dipolar
interaction is negligible compared with the $^{13}$C-$^{1}$H interaction.
However, when DD is active and removes the $^{13}$C-$^{1}$H interaction,
the $^{13}$C-$^{13}$C couplings become the dominant interaction
for causing decoherence. This holds strictly for ideal pulses, which
do not affect the $^{13}$C-$^{13}$C interaction. Under these conditions,
the decay rate is limited by the $^{13}$C-$^{13}$C couplings. However,
if pulses have finite duration, they can further extend the lifetime
of the component of the magnetization aligned with the effective field
\cite{Haeberlen1969,Rhim1976,Franzoni2005,li_generating_2007,Franzoni2008,li_intrinsic_2008,dong_controlling_2008}.

\section{Pulse Sequences}

\label{sec:pulses}

\subsection{Carr-Purcell (CP) and Carr-Purcell-Meiboom-Gill (CPMG)}

The Carr-Purcell (CP) sequence \cite{Carr1954} was first introduced
as a means to suppress inhomogeneity of the $B_{0}$ field $\Delta B_{0}^{j}$
seen at the $j^{\textrm{th}}$ nuclear site when the molecules containing
the spins are undergoing diffusion. The resulting time-dependent system-environment
interaction interferes with the refocusing process of a Hahn echo
\cite{Hahn1950} if the time dependence happens on a time scale comparable
to or faster than the refocusing time.

In our solid-state spin system, the time dependence of the environment
arises from fluctuating dipolar fields due to the neighboring spins
of the bath \cite{Haeberlen1976}. This generates a time-dependent
inhomogeneity. By suppressing it, the CP sequence acts as a method
of dynamical decoupling since, in effect, it suppresses the SE interaction
\cite{5916}.

The CP sequence consists of a train of refocusing pulses generating
spin-echoes. If the refocusing pulses are ideal, i.e., they cause
a complete $\pi$ rotation of the nuclear spins in an infinitesimal
time, and the spin-bath is static ($[\widehat{\mathcal{H}}_{E},\widehat{\mathcal{H}}_{SE}]=\hat{0}$),
then all terms $\widehat{\mathcal{H}}^{(m)}$ of the Magnus expansion
vanish, and the inhomogeneity is completely suppressed. However, this
does not happen when the spin-bath is not static or the pulses have
finite duration or contain errors. For ideal pulses but a fluctuating
environment only the average system environment Hamiltonian $\widehat{\mathcal{H}}_{SE}^{(0)}$,
the zeroth order of the Magnus expansion, is suppressed \cite{Alvarez2010}.

If flip-angle errors are also considered, the system-environment coupling
no longer vanishes, even in lowest order, $\widehat{\mathcal{H}}_{SE}^{(0)}\neq\hat{0}$,
and the resulting propagator increases with the flip-angle error.
Considering this, Meiboom and Gill suggested a modification to the
CP sequence \cite{Meiboom1958}, now called the CPMG sequence, where
the rotation axis of the pulses is the same as the orientation of
the initial state of the spins. In this case, flip-angle errors have
no effect in zeroth order. Fig. \ref{Flo:Pulse_Seq}a shows a CPMG
sequence. Note that the CPMG cycle requires only 2 pulses; hence a
sequence of 10 pulses covers 5 cycles. In the figure, $\tau$ is the
average distance between pulses, while $\tau_{c}$ denotes the length
of a single cycle.

\subsection{Uhrig dynamical decoupling (UDD)}

In the CPMG sequence, the separation between adjacent pulses is constant
throughout the sequence. In a seminal paper \cite{Uhrig2007,Uhrig2008},
Uhrig relaxed this condition, searching for the optimal combination
of delays that would minimize the effect of a purely dephasing system-environment
coupling. He showed that the optimal distribution for reducing low-frequency
noise corresponds to a sinusoidal modulation of the pulse delays.
More specifically, one cycle of an $N^{\textrm{th}}$ order UDD sequence
consists of $N$ pulses applied at times \begin{equation}
t_{i}=\tau_{c}\sin^{2}\left[\frac{\pi i}{2\left(N+1\right)}\right],\end{equation}
 where $t_{N+1}=\tau_{c}$ is the cycle time and $t_{0}=0$ the starting
time. Fig. \ref{Flo:Pulse_Seq}b shows the UDD sequence for $N=10$.
The CPMG sequence is the simplest UDD sequence with order $N=2$.
The UDD sequence was shown to be universal for {any} purely dephasing
Hamiltonian \cite{Yang2008}.

\section{Experiments}

\label{sec:expts}

\subsection{Measurement scheme}

\begin{figure}
\centering \includegraphics[scale=0.7]{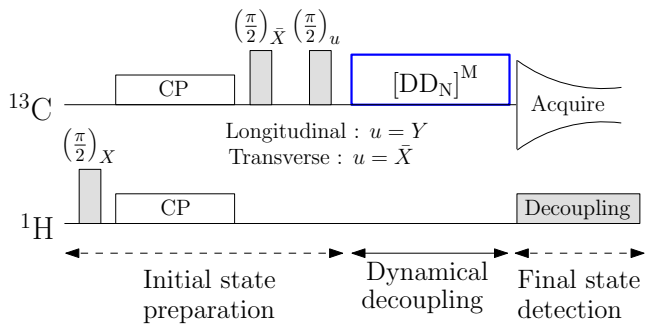}

\caption{(Color online) The experimental scheme: the initial state preparation
uses cross-polarization (CP) to enhance the polarization of the $^{13}$C
nuclear spins. The actual dynamical decoupling consists of $M$ cycles
of the $N$-order DD sequence. The DD pulses are applied along the
$X$ axis, while the initial states are created using a $\left(\frac{\pi}{2}\right)$pulse
with $Y$ (-$X$) phase for the longitudinal (transverse) initial
conditions.}

\label{fig:pulse-sequence} %
\end{figure}

\begin{figure}
\centering \includegraphics[scale=0.5]{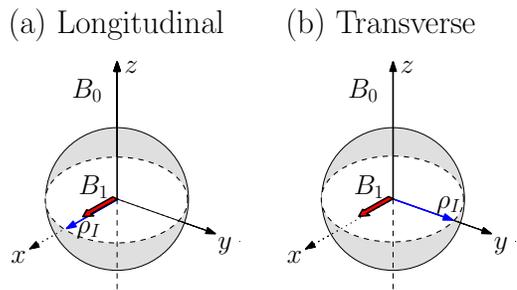}

\caption{(Color online) Different initial states of the spin system: The blue
(thin,long) arrows indicate the orientation of the initial $^{13}$C-spin
polarization, and the red (thick,short) arrows show the direction
in which the pulses are applied.}

\label{fig:initial-condition} %
\end{figure}

The general procedure used in our experiments is illustrated in Fig.
\ref{fig:pulse-sequence}. Cross-polarization from the abundant proton
spins to the $^{13}$C spins is used at the initial preparation stage
to increase the $^{13}$C polarization \cite{Pines72}. After this
transfer, we store the enhanced $^{13}$C polarization along the $z$
axis and wait for a time longer than the dephasing time to erase any
correlation that could arise during the cross-polarization process.
The magnetization is then rotated to an initial state, represented
by the density operator $\hat{\rho}_{0}=\hat{S}_{\{x,y\}}$, in the
$xy$-plane that is transverse to the strong static field. We use
two distinct initial states considering the relative phases of the
polarization of $\hat{\rho}_{0}$ and the pulses to be used for DD
\cite{Biercuk2009a,Alvarez2010}-- the {}``longitudinal'' state
in Fig. \ref{fig:initial-condition}(a) where the pulses are applied
in the direction of the $\hat{\rho}_{0}$ polarization, and the {}``transverse''
state in Fig. \ref{fig:initial-condition}(b) where they are applied
perpendicular to the $\hat{\rho}_{0}$ polarization. This translates,
in Fig. \ref{fig:pulse-sequence}, to applying the DD $\pi$ pulses
with an $X$ phase where a $\pi/2$ pulse is applied before of the
DD sequence with a $Y$ phase for the longitudinal initial condition,
and with a $-X$ phase for the transverse condition. $M$ cycles of
the $N$-pulse DD sequence are applied; the duration for the DD $\pi$
pulses was 10.4 $\mu$s. The errors of the pulse delays (compared
to the theoretical values) are 2.5 ns on average, limited by the timing
resolution of our pulse generator. The error of the pulse durations
(jitter) is < 0.1 ns. Therefore timing errors are negligible compared
to amplitude errors, which are of the order of 10\%. In general for
the range of delays between pulses used in our experiments, we did
not observe substantial differences between considering the delays
from the pulse edges or centers. This was also corroborated by numerical
simulations. After each DD cycle, we measured the remaining spin
polarization by recording the NMR signal \cite{Alvarez2010}. During
the signal acquisition, we applied continuous wave decoupling to prolong
the FID and thereby increase the detection sensitivity.

\begin{figure}
\includegraphics[scale=0.29]{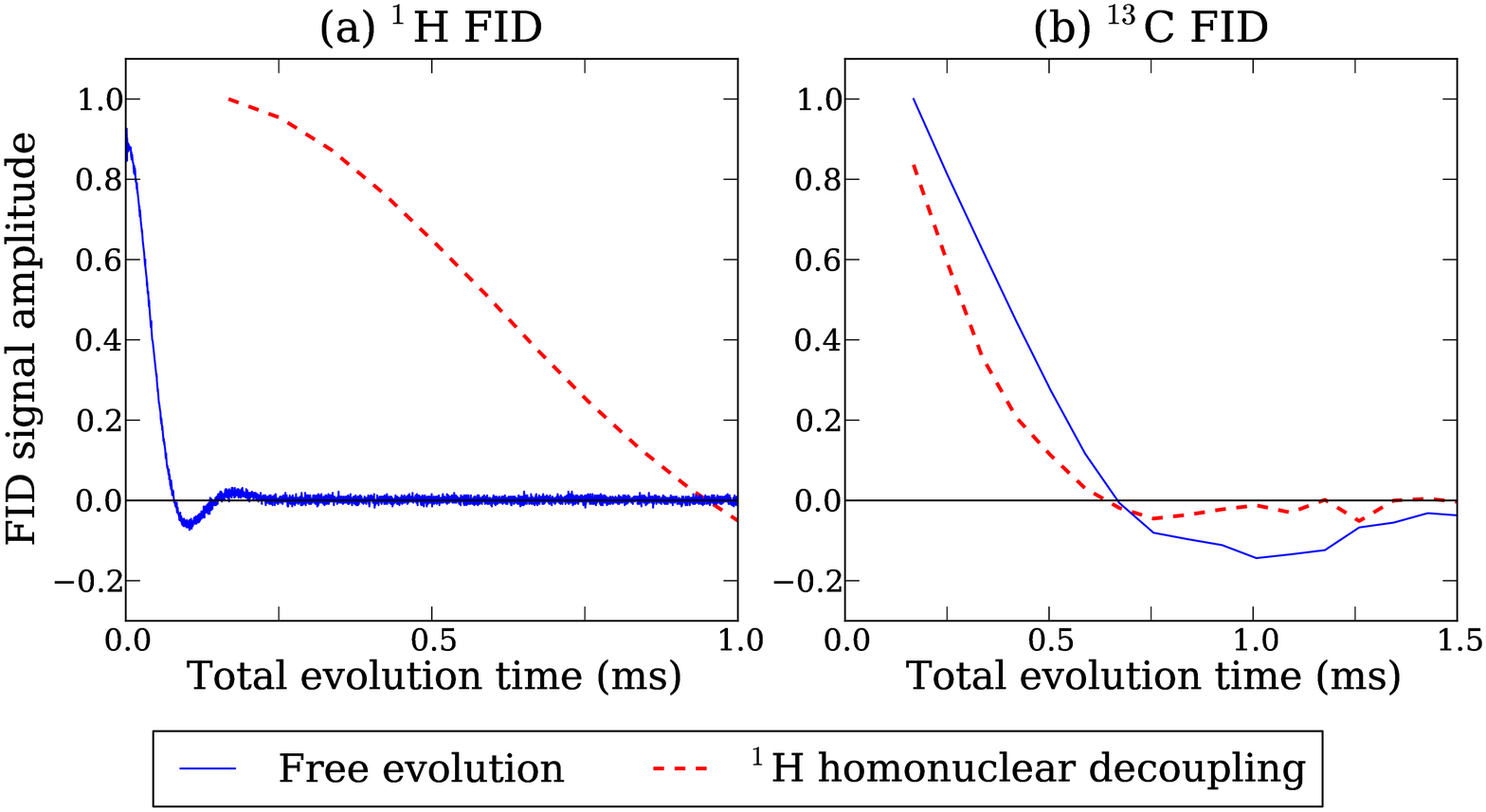}

\caption{(Color online) Experimental FID of (a) $^{1}$H and (b) $^{13}$C
spins without (blue solid line) and with (red dashed line) BLEW-12
$^{1}$H homonuclear decoupling. The correlation time of the bath
($^{1}$H) increases by an order of magnitude if the BLEW-12 sequence
is applied. The qubit system (the $^{13}$C nuclei) remains almost
unaffected by the BLEW-12 sequence.}

\label{fig:fidblew} %
\end{figure}

The decay of the signal amplitude as a function of the DD evolution
time $t=M\tau_{c}$ represents the survival probability of the state
$\hat{\rho}_{0}$: \[
s(t)=\frac{\mathrm{Tr}\left\{ \hat{\rho}_{0}\hat{\rho}\left(t\right)\right\} }{\mathrm{Tr}\left\{ \hat{\rho}_{0}\hat{\rho}_{0}\right\} },\]
 where $\hat{\rho}(t)=\hat{U}(t)\hat{\rho}_{0}\hat{U}^{\dagger}(t)$
is the density operator of the spin-system at time $t$. In the case
of free evolution (no DD), this corresponds to the free induction
decay signal represented in Fig. \ref{fig:fidblew}. The solid blue
lines in panels (a) and (b) represent the FID signal of the $^{1}$H
and $^{13}$C spins, respectively, which correspond to the observables
$\hat{I}_{x}$ and $\hat{S}_{x}$.

\begin{figure}
\centering \includegraphics[scale=0.3]{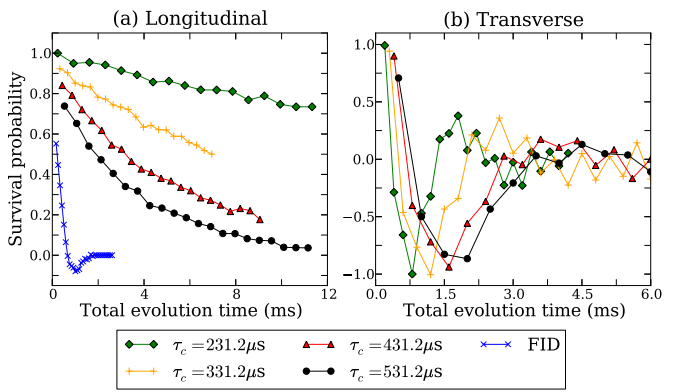}

\caption{(Color online) Time evolution of the $^{13}$C magnetization of the
initial state (proportional to its survival probability), for the
(a) longitudinal and (b) transverse initial conditions subjected to
UDD$_{3}$. The value of -1 in the latter corresponds to the state
that is antiparallel to the initial state.}

\label{fig:decaylong} %
\end{figure}

\subsection{Single parameter three pulse family}

\label{sec:expts1}

We first consider the simplest sequence that contains unequal spacings
between the pulses. Fig. \ref{fig:decaylong} (a) and (b) show the
time evolution of the $^{13}$C magnetization from a longitudinal
initial state $\hat{\rho}_{0}=\hat{S}_{x}$ and a transverse initial
state $\hat{\rho}_{0}=\hat{S}_{y}$ for different cycle times. They
show that when the longitudinal case is compared to the FID, dynamical
decoupling allows for a longer survival of coherence, and this improves
for shorter cycle times \cite{Alvarez2010}. The signal decay is an
order of magnitude faster for the transverse initial condition in
Fig. \ref{fig:decaylong}(b), than the longitudinal condition \ref{fig:decaylong}(a)
\cite{Alvarez2010}. As we demonstrated in Ref. \cite{Alvarez2010},
in this case the decay is due mostly to pulse imperfections (flip-angle
errors), which lead to a loss of polarization. In the longitudinal
case, the effect of these pulse errors is partly compensated over
the sequence, but in the transverse case, they accumulate. This accounts
for the large difference between the two cases already demonstrated
in previous works \cite{Biercuk2009,Alvarez2010}. In particular in
our experimental setup, the flip-angle error was estimated to be around
10\% \cite{Alvarez2010}.

\begin{figure}
\includegraphics[scale=0.29]{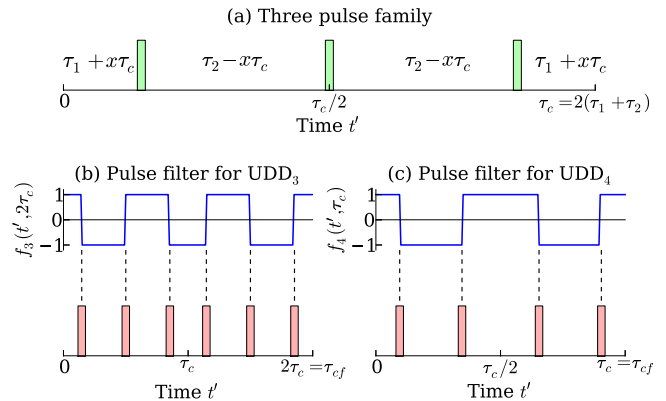}

\caption{(Color online) (a) Single parameter sequence family containing all
mirror symmetric three-pulse sequences. They are described by the
parameter $x$, where $\tau_{1}$ and $\tau_{2}$ are the exact UDD$_{3}$
delays. (b) Time domain filter function $f_{3}(\tau^{\prime},2\tau_{c})$
for two cycles of the UDD$_{3}$ sequence. (c) Time domain filter
function $f_{4}(\tau^{\prime},\tau_{c})$ for one cycle of the UDD$_{4}$
sequence. In the filter representation we assume perfect pulses in
the center of the finite-width pulses. The dashed lines indicate that
the time domain filter toggles between $\pm1$ at the position of
the pulses.}

\label{fig:three-pulse} %
\end{figure}

Uhrig introduced the idea that it might be advantageous to omit the
condition that all pulse separations should be identical and derived
a simple formula that determines the optimal spacing of $N$ pulses.
We examine this scheme first in the simplest possible case, corresponding
to a cycle of 3 pulses. In order to determine the optimal pulse spacing
experimentally, we parametrize the pulse spacings of the symmetric
three-pulse sequence with a parameter $x,$ as shown in Fig. \ref{fig:three-pulse}(a).
$x$ is defined as the deviation (as a fraction of the cycle time)
of the pulse separation from those of the UDD$_{3}$ sequence, which
are $\tau_{1}=\tau_{c}\sin^{2}(\pi/8)$ and $\tau_{2}=\tau_{c}/2-\tau_{1}$.
This parametrization captures all possible mirror symmetric three-pulse
sequences. When $x=0$, the sequence is UDD$_{3}$, and when $x\simeq0,0203$,
the sequence is CPMG. For this family of DD sequences, we experimentally
determined the decay rate as a function of $x$ by fitting a pure
exponential function.

\begin{figure}
\includegraphics[scale=0.32]{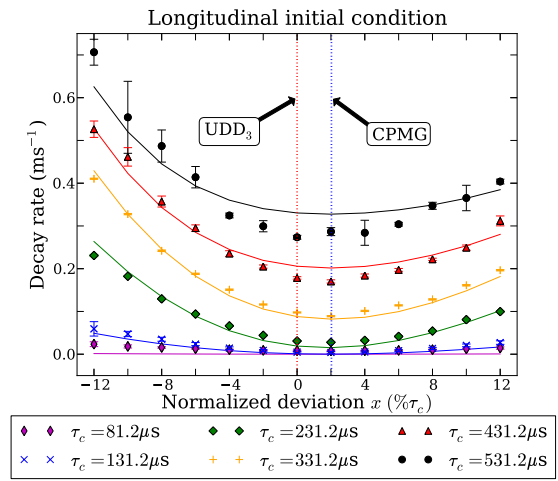} 

\caption{(Color online) Experimental (symbols) and simulated (lines) decay
rates of the magnetization as a function of the deviation $x$ for
different cycle times for the longitudinal initial condition. The
UDD$_{3}$ sequence corresponds to $x_{{\rm {\mathrm{UDD}}}}=0$,
while the CPMG sequence corresponds to $x_{{\rm {\mathrm{CPMG}}}}\approx0.0203.$
Note that the quadratic fitting curves used to determine to position
of the minimum of each curve are not shown.}

\label{fig:devmain} %
\end{figure}

\begin{figure}
\includegraphics[scale=0.32]{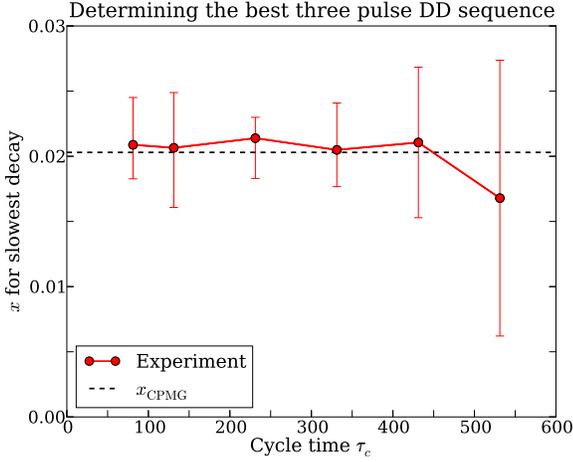} 

\caption{(Color online) In order to determine the best longitudinal three-pulse
DD sequence, the decay curves in Fig. \ref{fig:devmain} were fitted
to a quadratic function in the region of the minimum. The positions
of the minima of these quadratic fits are represented by the red circles
in this figure. The black dotted line corresponds to $x_{{\rm {\mathrm{CPMG}}}}\approx0.0203.$}

\label{fig:quadfit} %
\end{figure}

Figure \ref{fig:devmain} shows the experimentally obtained decay
rates for the longitudinal initial condition, as a function of the
parameter $x$. We determined the value of $x$ that generates the
slowest decay rate by fitting the experimental data points with a
quadratic function (not shown in the figure) and determining the
position of the minimum. These minima are the red circles in Fig.
\ref{fig:quadfit}. The sequence that generates the slowest decay
is obtained for $x_{min}\approx0.021\pm0.002$, for all cycle times
used in the experiment. This is astonishingly close to the value $x_{\mathrm{CPMG}}=0.0203$
that corresponds to the equally spaced sequence.

In addition to the experimental data points, Fig. \ref{fig:devmain}
also shows solid lines. They represent the results of numerical simulations
of the decay rates for the same parameters as the experiments. In
the simulations, we assumed that the pulses were ideal. Following
Ref. \cite{Cywinski2008,Uhrig2008} this involves modeling the pulse
sequence as a one-dimensional filter $f(\tau,\tau_{c})$, as shown
in \ref{fig:three-pulse}(b). A detailed description of this model
and the simulations is presented in Sec. \ref{sec:physics}. Since
this model does not take into account pulse errors, it cannot be applied
to the transverse initial condition, where these errors play a predominant
role \cite{Alvarez2010}.

For the transverse initial condition, the rates are about an order
of magnitude larger than for the longitudinal initial condition, and
they increase for short cycle times \cite{Alvarez2010}. They are
essentially independent of the parameter $x$ because they are dominated
by pulse errors. We therefore consider only longitudinal states in
the following.

%


%

\subsection{Variation of UDD order}

\label{sec:expts2}The performance of a dynamical decoupling sequence
increases with the average {power} of the external control field
used for DD \cite{5916}. This means that given a fixed time window
$t$, the DD efficiency improves with the number of refocusing pulses.
Since the pulses are of finite length, the maximum number of pulses
that can be employed during $t$ is bounded \cite{viola2003,Hodgson2010,Uhrig2010,UhrigLidar2010,Alvarez2010,viola2010}.

\begin{figure}
\centering \subfigure{\includegraphics[scale=0.29]{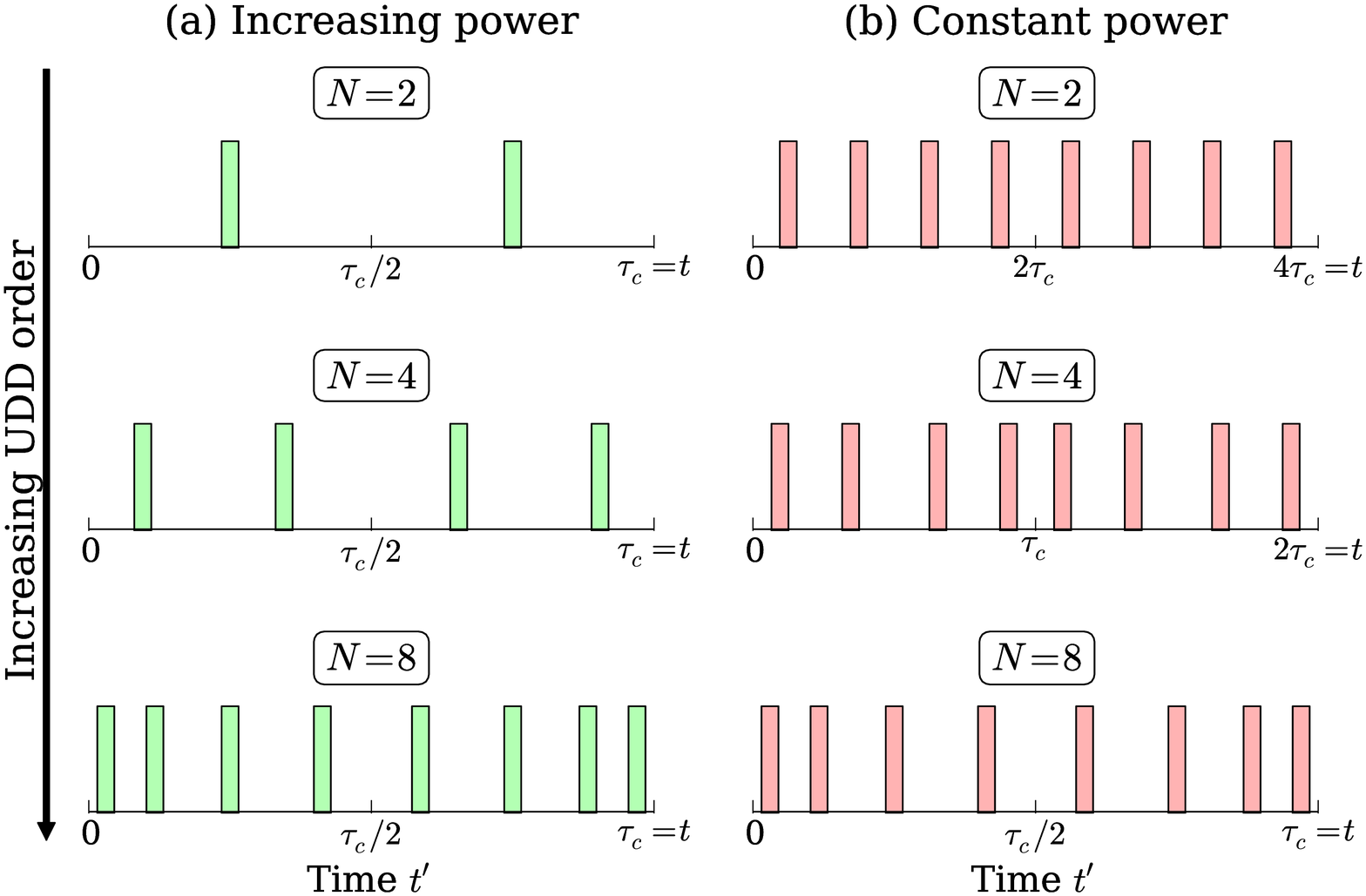}} \subfigure{\includegraphics[scale=0.29]{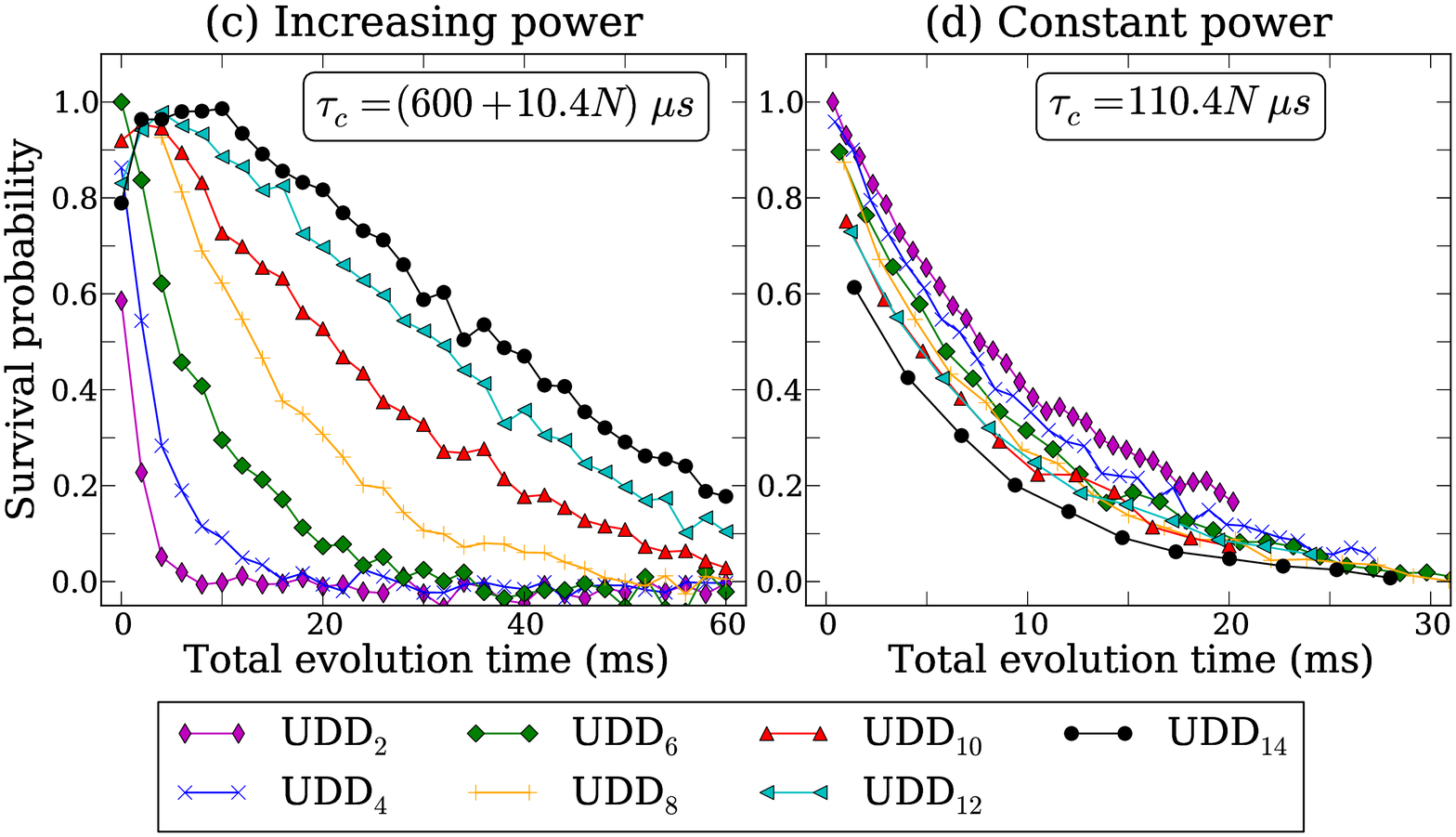}}

\caption{(Color online) Top panels: Pulse sequence scheme keeping fixed the
time window $t$ as the UDD order $N$ is increased. To keep the DD
power constant across UDD orders, the cycle time is scaled with $N$
(b). Bottom panels: Experimental $^{13}$C signal decay for different
UDD orders for (c) a fixed cycle time $\tau_{c}=600\mu$s, (d) a scaled
cycle time $\tau_{c}=110.4N\mu$s.}

\label{fig:power} %
\end{figure}

The scheme depicted in Fig. \ref{fig:power}(a) describes a situation
in which the cycle time $\tau_{c}$ is fixed, while the UDD order
is increased from top to bottom. Given the fixed window $t=\tau_{c}$,
the power used for DD during $t$ increases from top to bottom. The
corresponding survival probability is shown in Fig. \ref{fig:power}(c)
for different UDD orders (from UDD$_{2}$ to UDD$_{14}$) and with
$\tau_{c}=600\mu s$ (not including the pulse lengths). Clearly, if
the cycle time is fixed, higher order UDD sequences maintain the qubit
coherence for longer times. However, this improved performance is
obtained at the price of an increase in the average power applied
to the system. For many applications, this is not possible, and a
more meaningful comparison is obtained by keeping the average power
level fixed. Fig. \ref{fig:power}(b), shows the corresponding pulse
sequences and Fig. \ref{fig:power}(d) the resulting decays. For these
experiments, the cycle time was scaled with the UDD order $N$ as
$\tau_{c}=N\,110.4\,\mu$s. Under these conditions, the variation
of the decay rate with the UDD order is relatively weak and higher
orders perform slightly worse than UDD$_{2}$(CPMG) .

\begin{figure}
\includegraphics[scale=0.29]{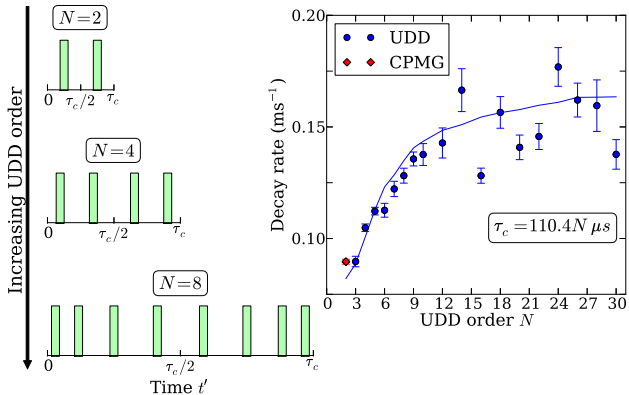}

\caption{(Color online) Left panel: Pulse scheme manifesting the cycle time
scaling with the UDD order $N$. Right panel: Experimental (symbols)
and simulated (line) decay rates of the $^{13}$C magnetization for
different UDD orders (blue circles) compared with decay rates achieved
with the CPMG sequence (red rhombus). The same number of pulses during
a time window $t$ are applied by scaling the cycle time as $\tau_{c}=110.4N\,\mu$s. }

\label{fig:ordermain} %
\end{figure}

We quantified the decay rates for UDD orders 2 to 30 for $\tau_{c}=N\,110.4\,\mu$s.
The results are shown in the right panel of Fig. \ref{fig:ordermain}.
The left panel schematizes how the DD power is kept constant across
orders -- the cycle time is scaled with the UDD order. Clearly, for
a large span of UDD orders, the CPMG sequence performs better than
the UDD. The results mean two things: first, that in the regime of
our experiment, higher UDD orders provide no advantage over lower
UDD orders; and second, that in consequence the UDD protocol itself
performs worse than the CPMG sequence with the same number of pulses.
Although these experiments were carried out with the average delay
between pulses of $110.4\mu$s, the same qualitative behavior is also
observed for different times between pulses. However, the difference
between the decay rates of CPMG and UDD can be reduced by shortening
the cycle time as we show below. %

%

\subsection{Variation of cycle time}

\label{sec:expts3} 
\begin{figure}
\includegraphics[scale=0.29]{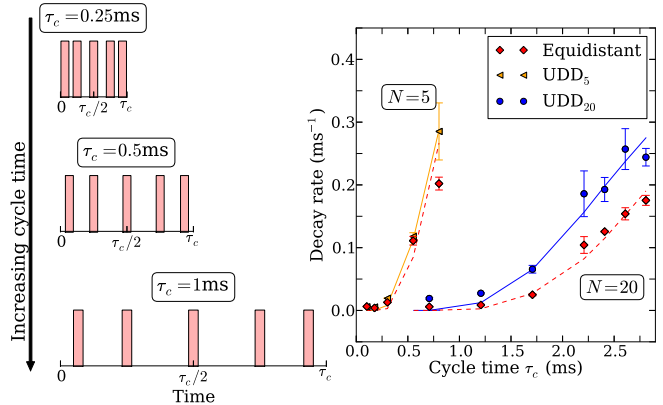}

\caption{(Color online) Left panel: Pulse sequence scheme showing the variation
of its cycle time for a fixed UDD order $N=5$. Right panel: Experimental
(points) and simulated (lines) decay rates of the $^{13}$C magnetization
for UDD$_{5}$ and UDD$_{20}$ and its comparison with equidistant
pulse (CPMG) sequence having the same number of pulses for the different
UDD cycle times. }

\label{fig:time} %
\end{figure}

In the limit of infinitesimally short cycle time
and infinitely strong pulses -- the \textit{bang-bang} regime \cite{5916}--
one can maintain quantum coherence for arbitrarily long times. However,
in any real experiment, the peak power as well as the average power
applied to the system are limited and should be minimized. Accordingly,
the duration of the pulses and the cycle time cannot be reduced below
some minimal values \cite{viola2003,Hodgson2010,UhrigLidar2010,Alvarez2010}.
In this section, we examine the DD performance of a given sequence
as a function of the cycle time. The left panel of Fig. \ref{fig:time}
schematically describes the variation of the cycle time for a fixed
UDD order $N$, in this case for $N=5$. In the right panel of Fig.
\ref{fig:time}, the points are experimentally obtained decay rates
while the lines are simulation results using the filter model for
ideal pulses detailed in Sec. \ref{sec:physics}. For long cycle times,
the performance of equidistant pulses (CPMG) is better than UDD however
as $\tau_{c}$ is reduced its performance improves considerably, although
the UDD decay rates are indistinguishable from the CPMG in this regime.
Similar results were obtained for similar experiments with different
UDD orders. Fig. \ref{fig:time} shows that in the regime of our experimental
setup, UDD does not perform better than the CPMG for any $\tau_{c}$.

%

\subsection{Reducing the fluctuation rate}

\label{sec:expts4} The decay rate of the survival probability under
dynamical decoupling depends on the ratio of the cycle time $\tau_{c}$
to the correlation time of the bath $\tau_{B}$ \cite{khodjasteh_performance_2007,Alvarez2010}.
The UDD sequence was designed to provide optimal decoupling performance
for baths with long correlation times. In our system, the bath correlation
time depends on the strength of the dipolar couplings $d_{ij}$ between
the $^{1}$H nuclear spins of the bath \cite{Abragam}. One can effectively
rescale $d_{ij}$, and hence the correlation time of the bath, by
applying decoupling sequences to the bath spins \cite{Haeberlen1976}.
This allows us to access regimes where the cycle time $\tau_{c}$
is significantly shorter than the correlation time of the bath.. For
this we employ the well-known homonuclear decoupling sequence BLEW-12
\cite{burum1981}. As shown by the dashed lines of Fig. \ref{fig:fidblew},
application of this pulse sequence increases the bath-correlation
time by about one order of magnitude: it was around $\tau_{B}\approx100\mu$s
without BLEW-12 and $\tau_{B}\approx1000\mu$s with BLEW-12 for a
cycle time $\tau_{cB}$ = $84\mu$s. To determine the effects on the
qubit system, we measured its free induction decay (FID). We found
that it remains almost unaffected by the $^{1}$H homonuclear decoupling
(Fig. \ref{fig:fidblew}b). The scaling due to the BLEW sequence \cite{burum1981},
0.475, appears to be indistinguishable from the effect of self-decoupling,
which reduces the $^{13}$C linewidth in the absence of $^{1}$H-homonuclear
decoupling \cite{Ernst1998}. The experimental scheme of Fig. \ref{Flo:Pulse_Seq}
was used again, the only change being that the homonuclear decoupling
sequence is applied to the bath in parallel with the application of
the DD sequence to the system.

\begin{figure}
\includegraphics[scale=0.29]{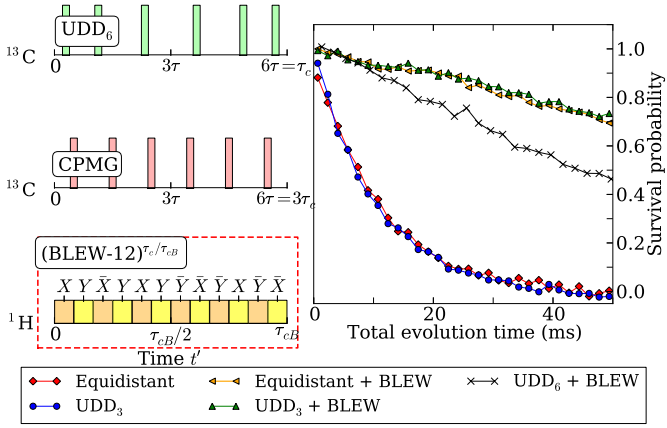}

\caption{(Color online) Left panel: DD pulse sequence scheme for UDD$_{6}$
and CPMG applied in conjunction with the BLEW-12 sequence on the $^{1}$H
spin-bath. The cycle time of the DD sequences is $6\tau$, while for
the BLEW-12 sequence the cycle time is $\tau_{cB}$, where $6\tau/\tau_{cB}$
is an integer. Right panel: $^{13}$C signal decay for the DD sequences
described in the legend. }

\label{fig:blew} %
\end{figure}

Figure \ref{fig:blew} shows the pulse sequence used during the DD
period (left panel). For different UDD orders, the signal decays were
measured with and without the BLEW-12 sequence (dashed box in the
left panel) on the $^{1}$H channel. The experimental results in the
right panel of Fig. \ref{fig:blew} show that the increase of the
bath correlation time by roughly one order of magnitude leads to significant
improvements in the performance of all DD sequences. The performance
of UDD$_{3}$ and CPMG is indistinguishable, while that of UDD$_{6}$
is lower.

\begin{figure}
\centering \subfigure{\includegraphics[scale=0.29]{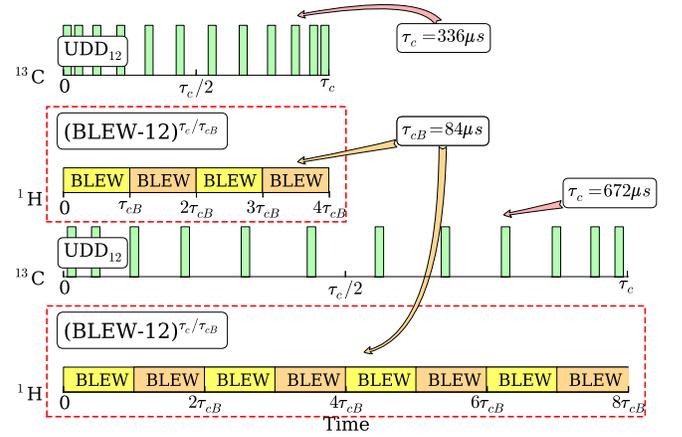}} \subfigure{\includegraphics[scale=0.37]{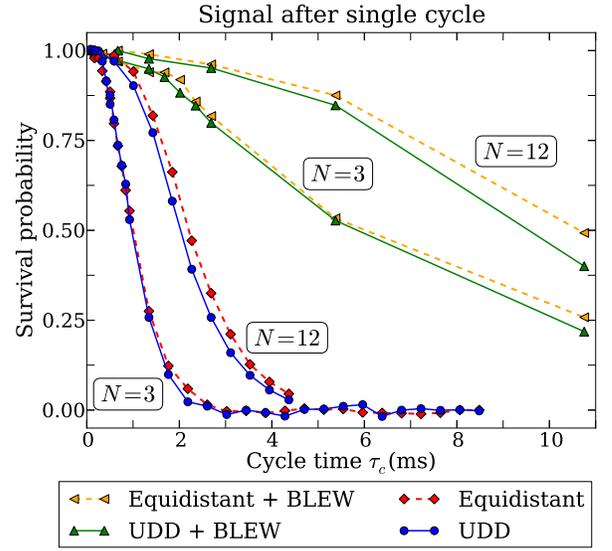}}

\caption{(Color online) (a) UDD$_{12}$ sequence applied with homonuclear decoupling
of the $^{1}$H spins. Sequences for two different cycle times $\tau_{c}=336\mu$s
= $4\tau_{cB}$ and $\tau_{c}=672\mu$s $=8\tau_{cB}$, where $\tau_{cB}=84\mu$s
is the cycle time of the BLEW-12 homonuclear decoupling sequence.
(b) Experimental survival probability (fidelity) of the initial state
$\hat{\rho}_{I}=\hat{S}_{x}$ for a single cycle with $N$ pulses
and varying the cycle time $\tau_{c}$ with and without homonuclear
decoupling of the spin-bath.}

\label{fig:blew2} %
\end{figure}

Similar results are obtained for longer and shorter cycle times. This
is evidenced in Fig. \ref{fig:blew2}(b), which shows the survival
probability after a \textit{single} cycle of the UDD and equidistant
pulse (CPMG) sequences for different cycle times and DD orders. The
pulse sequence for this experiment is shown in Fig. \ref{fig:blew2}(a)
for two cycle times $\tau_{c}=336\mu$s and $\tau_{c}=672\mu$s of
the UDD$_{12}$ sequence. The results show that for large UDD orders
like $N=12$, the CPMG sequence performs better, even when the fluctuations
of the environment are slow. For $N=3$ the performance of both sequences
is comparable within experimental errors.

%

\section{Physical interpretation of the experimental results}

\label{sec:physics}

\subsection{Semiclassical approximation\label{sub:Semi-classical-Aapproximation}}

In this section, we present a semiclassical model to explain and interpret
the experimental results of section \ref{sec:expts}. Since the system
is well in the high-temperature limit, it is possible to use a semiclassical
description instead of the fully quantum mechanical treatment \cite{Abragam}.
Most of the mathematical description of this subsection was developed
within the DD context on different systems \cite{Cywinski2008,Uhrig2008}
and they can be obtained from standard semiclassical treatments \cite{Abragam}.
Here we connect and reinterpret them to describe our spin system.
Starting from the quantum mechanical description of section \ref{sec:system-2},
the effect of the environment-Hamiltonian $\mathcal{\widehat{H}}_{E}$
on the evolution of the system may be discussed in an interaction
representation with respect to the evolution of the isolated environment:
the system-environment Hamiltonian then becomes\begin{align}
\mathcal{\widehat{H}}_{SE}^{(E)}\left(t\right) & =e^{-i\mathcal{\widehat{H}}_{E}t}\mathcal{\widehat{H}}_{SE}e^{i\mathcal{\widehat{H}}_{E}t}\nonumber \\
 & =\hat{S}_{z}e^{-i\mathcal{\widehat{H}}_{E}t}\left(\sum_{j}b_{Sj}\hat{I}_{z}^{j}\right)e^{i\mathcal{\widehat{H}}_{E}t}\label{eq:HSE-time-dependent}\\
 & =b_{SE}\hat{S}_{z}e^{-i\mathcal{\widehat{H}}_{E}t}\hat{E}_{z}e^{i\mathcal{\widehat{H}}_{E}t},\end{align}
 where $\hat{E}_{z}=\left[\sum_{j}\frac{b_{Sj}}{b_{SE}}\hat{I}_{z}^{j}\right]$
represents an effective spin bath operator and $b_{SE}=\sqrt{\sum_{i}b_{Si}^{2}}$
the coupling strength. Since $\mathcal{\widehat{H}}_{E}$ does not
commute with $\mathcal{\widehat{H}}_{SE}$, the effective system-environment
interaction $\mathcal{\widehat{H}}_{SE}^{\left(E\right)}$ becomes
time-dependent: the system experiences a coupling to the environment
that fluctuates. For a semi-classical treatment we trace over the
bath variables, and replace $b_{SE}e^{-i\mathcal{\widehat{H}}_{E}t}\hat{E}_{z}e^{i\mathcal{\widehat{H}}_{E}t}$
of Eq. (\ref{eq:HSE}) by the stochastic function $b_{SE}E_{z}(t)$
representing a classical random field with a Gaussian distribution
with zero average, $\left\langle E_{z}(t)\right\rangle =0$, and the
autocorrelation function $\left\langle E_{z}(t)E_{z}(t+\tau)\right\rangle =g\left(\tau\right)$
\cite{Abragam}. The spectral density of the system-bath interaction
is the Fourier transform of $g(\tau)$, \begin{equation}
S(\omega)=\frac{1}{\sqrt{2\pi}}\int_{-\infty}^{\infty}d\tau\, g(\tau)\, e^{-i\omega\tau}.\label{WK}\end{equation}
 It describes the relative weight of the different frequency components
of the SE interaction.

Using this effective field $E_{z}(t)$, we write the semiclassical
SE interaction Hamiltonian as \begin{equation}
\widehat{\mathcal{H}}_{SE}(t)=b_{SE}E_{z}(t)\hat{S}_{z}.\label{eq:HSE2}\end{equation}
 Clearly, $\widehat{\mathcal{H}}_{SE}(t)$ commutes with itself at
all times. This allows us to calculate the survival probability \begin{equation}
s(t)=\frac{\textrm{Tr}\left\{ e^{-i\int_{0}^{t}dt_{1}\widehat{\mathcal{H}}_{SE}(t_{1})}\hat{\rho}_{0}e^{i\int_{0}^{t}dt_{1}\widehat{\mathcal{H}}_{SE}(t_{1})}\hat{\rho}_{0}\right\} }{\textrm{Tr}\left\{ \hat{\rho}_{0}\hat{\rho}_{0}\right\} }\end{equation}
 by integrating\begin{equation}
\int_{0}^{t}dt_{1}\widehat{\mathcal{H}}_{SE}(t_{1})=b_{SE}\hat{S}_{z}\int_{0}^{t}dt_{1}E_{z}(t_{1})=\phi(t)\hat{S}_{z}.\end{equation}
 The survival probability then becomes \begin{equation}
s(t)=\frac{\textrm{Tr}\left\{ e^{-i\phi(t)\hat{S}_{z}}\hat{\rho}_{0}e^{i\phi(t)\hat{S}_{z}}\hat{\rho}_{0}\right\} }{\textrm{Tr}\left\{ \hat{\rho}_{0}\hat{\rho}_{0}\right\} }.\end{equation}
 If the spin ensemble is initially polarized along the $z$-direction,
$\hat{\rho}_{0}=\hat{S}_{z}$, $\hat{S}_{z}$ is a constant of motion.
However if $\hat{\rho}_{0}=\hat{S}_{x,y}$, its survival probability
is\begin{align}
s_{x,y}(t) & =\frac{\textrm{Tr}\left\{ e^{-i\phi(t)\hat{S}_{Z}}\hat{S}_{x,y}e^{i\phi(t)\hat{S}_{Z}}\hat{S}_{x,y}\right\} }{\textrm{Tr}\left\{ \hat{S}_{x,y}^{2}\right\} }=\cos\phi(t).\end{align}
 Taking the average over the random fluctuations \begin{equation}
\left\langle s_{x,y}(t)\right\rangle =\left\langle \cos\phi(t)\right\rangle =e^{-\frac{1}{2}\left\langle \phi^{2}(t)\right\rangle },\end{equation}
 where we have used $\cos\phi(t)=\left(e^{i\phi(t)}+e^{-i\phi(t)}\right)/2$
and the property $\left\langle e^{iX}\right\rangle =e^{i\left\langle X\right\rangle -\left\langle X^{2}\right\rangle /2}$
for a Gaussian random variable $X$. For a simple interpretation of
the decay, we use the exponential's argument\begin{multline}
\chi(t)=\frac{1}{2}\left\langle \phi^{2}(t)\right\rangle =\frac{|b_{SE}|^{2}}{2}\left\langle \int_{0}^{t}dt_{1}\int_{0}^{t}dt_{2}E_{z}(t_{1})E_{z}(t_{2})\right\rangle \\
=\frac{|b_{SE}|^{2}}{2}\int_{-\infty}^{\infty}dt_{1}\int_{-\infty}^{\infty}dt_{2}\: g(t_{1}-t_{2})f_{0}(t_{1},t)f_{0}(t_{2},t),\label{eq:convolution}\end{multline}
 where $f_{0}(t^{\prime},t)=\Theta(t^{\prime})\Theta(t-t^{\prime})$
with the Heaviside function $\Theta(t^{\prime})=1\:\forall\:\: t^{\prime}>0$
and zero otherwise. Using Eq. (\ref{WK}), we rewrite this convolution
integral as \begin{equation}
\chi(t)=\frac{\sqrt{2\pi}|b_{SE}|^{2}}{2}\int_{-\infty}^{\infty}d\omega S(\omega)\left|F_{0}(\omega,t)\right|^{2},\label{eq:decaychi}\end{equation}
 where $F_{0}(\omega,t)$ is the Fourier transform of $f_{0}(t^{\prime},t)$
\cite{Cywinski2008} \begin{gather}
F_{0}(\omega,t)=\frac{1}{\sqrt{2\pi}}\int_{-\infty}^{\infty}dt^{\prime}f_{0}(t^{\prime},t)e^{-i\omega t^{\prime}}\nonumber \\
=\frac{1}{\sqrt{2\pi}}\int_{0}^{t}dt^{\prime}e^{-i\omega t^{\prime}}=\frac{1}{\sqrt{2\pi}}e^{-i\omega t/2}\frac{\sin\left(\omega t/2\right)}{\left(\omega/2\right)}\label{sumovertime}\end{gather}
 and we have used the fact that the random field $E_{z}$ is not correlated
with the time-domain filter function $f$. The decay function $\chi(t)$
is thus equal to the product of the spectral density $S(\omega)$
of the system-environment coupling and the filter transfer function
$F_{0}(\omega,t)$.

Here, we have treated the case where no control pulses are applied,
which corresponds to the free induction decay. Since both $S(\omega)$
and $\left|F_{0}(\omega,t)\right|^{2}$ reach their maximum at $\omega=0$,
it is the low frequency environmental noise that has the highest contribution
to the decay rate.

\subsection{Analogy between FID and single slit diffraction}

Since the decay function $\chi(t)$ arises from the interference between
the random fields $E_{z}(t_{1}),E_{z}(t_{2})$ at different times,
it is helpful to draw an analogy with interference effects in optics.
This analogy is best seen with the help of Huygens' principle, which
allows us to associate an elementary wave with every point in space;
here, we observe interference between elementary waves generated at
different points in time and weighted by the filter function $F(\omega,t).$
In the case of free precession (the FID), the time-domain filter function
$f_{0}(t^{\prime},t)$ is constant over the interval $t^{\prime}=[0,t]$.
This is exactly analogous to the case of diffraction from a slit that
extends from $0$ to $t$, and as in the optical case, we obtain a
diffraction pattern $\propto|\sin(x)/x|^{2}$.

According to Eq. (\ref{sumovertime}), the width of the diffraction
pattern is $\propto\frac{1}{t}$ and its amplitude $\propto t$. For
very short times, i.e., narrow slits, the corresponding diffraction
pattern is broader than the width of the spectral density $S(\omega)$
and the integral (\ref{eq:decaychi}) grows $\sim t^{2}$ . When the
time $t$ exceeds the correlation time $\tau_{B}$ of the system-environment
interaction, the width of slit broadens and the width of the filter
function $F_{0}(\omega,t)$ becomes narrower than the spectral density
pattern $S(\omega)$. In the long time limit $t\gg\tau_{B}$, the
filter function narrows to a delta function at $\omega=0$ and the
decay function becomes $\chi(t)\propto|b_{SE}|^{2}S(0)t$, corresponding
to an exponential decay. This result is equivalent to the one obtained
by Fermi's golden rule and is valid until a power law decay arises
\cite{Elena06,Alvarez2010Mesos}.

\subsection{Effect of pulses : interference}

The effect of DD pulses is a modulation of the time-domain filter
function $f(t',t)$ and therefore of the transfer or filter function
$F(\omega,t).$ As shown by Eq (\ref{eq:decaychi}), a slow-down of
the decay is achieved by minimizing the overlap between $S(\omega)$
and $\left|F(\omega,t)\right|^{2}$ \cite{Cywinski2008,Uhrig2008,Gordon2008,Biercuk2009a,Uys2009,Clausen2010,pan2010}.
In the typical case that the environmental spectral density peaks
at small frequencies, this implies that $\left|F(\omega,t)\right|^{2}$
should be close to $0$ for small frequencies $\omega$.

Let us now study the effect of ideal DD pulses. They generate reversals
of $\widehat{\mathcal{H}}_{SE}(t)$, so that the resulting Hamiltonian
still commutes with itself at all times. As described in the previous
sections, $N$ pulses are applied during the interval $\tau_{c}$
at positions $t_{j}=\{t_{1},t_{2},\cdots,t_{N}\}$, with $t_{0}=0$
and $t_{N+1}=\tau_{c}$. Under this condition, the time-domain filter
function $f_{N}(\tau^{\prime},\tau_{c})$ in Eq. (\ref{eq:convolution})
becomes \begin{equation}
f_{N}(t^{\prime},\tau_{c})=\sum_{j=0}^{N}\left(-1\right)^{j}\Theta(t^{\prime}-t_{j})\Theta(t_{j+1}-t^{\prime}),\end{equation}
 where $f_{N}(t^{\prime},\tau_{c})$ now switches between $\pm1$
at the position of every pulse \cite{Cywinski2008}. This is depicted
in Fig. \ref{fig:three-pulse}(c) for a cycle of UDD$_{4}$. We define
$\tau_{cf}$ as the period of the filter function $f_{N}(t^{\prime},\tau)$.
For even DD orders, $\tau_{cf}$ is equal to the cycle time $\tau_{c}$,
for odd DD orders, it equals {two} cycles, i.e., $\tau_{cf}=2\tau_{c}$
{[}Fig. \ref{fig:three-pulse}(b){]}.

If the average of the function $f_{N}(t^{\prime},\tau_{cf})$ over
the period $\tau_{cf}$ vanishes, the filter function vanishes at
$\omega=0$ \begin{equation}
\left|F_{N}(0,\tau_{cf})\right|=\frac{1}{\sqrt{2\pi}}\left|\int_{-\infty}^{\infty}dt^{\prime}f_{N}(t^{\prime},\tau_{cf})\right|^{2}=0.\end{equation}
 For a general $N$-pulse sequence, the transfer function $F_{N}(\omega,t)$
becomes \cite{Cywinski2008}\begin{multline}
F_{N}(\omega,\tau_{cf})=\frac{1}{\sqrt{2\pi}}\int_{0}^{\tau_{cf}}f_{N}(t^{\prime},\tau_{cf})e^{-i\omega t^{\prime}}dt^{\prime}\\
=\frac{1}{\sqrt{2\pi}}\sum_{j=0}^{N}\left(-1\right)^{j}\int_{t_{j}}^{t_{j+1}}e^{-i\omega t^{\prime}}dt^{\prime}\\
=\frac{1}{\sqrt{2\pi}}\frac{1+(-1)^{N+1}e^{-i\omega\tau_{cf}}+2\sum_{j=1}^{N}(-1)^{j}e^{-i\omega t_{j}}}{i\omega}.\label{generaln}\end{multline}

Uhrig found a suitable distribution of $N$ pulses over a time $\tau_{cf}$
that eliminates the first $N$ derivatives of $\omega F_{N}(\omega,\tau_{cf})$
\cite{Uhrig2007,Uhrig2008} and thus it has an optimally flat stop-band
at $\omega=0$.

For a given distribution of pulses, the corresponding time domain
filter function is \begin{equation}
f_{N}(t^{\prime},\tau_{cf})=f_{N}(t^{\prime},\infty)\times\Theta(t^{\prime})\Theta(\tau_{cf}-t^{\prime}),\end{equation}
 where $f_{N}(t^{\prime},\infty)$ is an infinite extension of the
filter function. $f_{N}(t^{\prime},\infty)$ can be written as a Fourier
series\begin{equation}
f_{N}(t^{\prime},\infty)=\sum_{k=-\infty}^{\infty}A_{k}\exp(ik\omega_{0}t^{\prime}),\label{eq:FTdecompositionf}\end{equation}
 where $k\omega_{0}=2\pi k/\tau_{cf}$ are the \textit{harmonic} frequencies
of the period $\tau_{cf}$, and \begin{align}
A_{k} & =\frac{\sqrt{2\pi}}{\tau_{cf}}F_{N}\left(k\omega_{0},\tau_{cf}\right)\end{align}
 the amplitudes. Hence, by convolution, the frequency domain filter
has the form, \begin{multline}
F_{N}(\omega,\tau_{cf})=\\
\sum_{k=-\infty}^{\infty}F_{N}\left(k\omega_{0},\tau_{cf}\right)e^{-i(\omega-k\omega_{0})\tau_{cf}/2}\frac{\sin\left[(\omega-k\omega_{0})\tau_{cf}/2\right]}{\left(\omega-k\omega_{0}\right)\tau_{cf}/2}.\label{eq:FTfrequenydomain}\end{multline}

\begin{figure}
\includegraphics[scale=0.37]{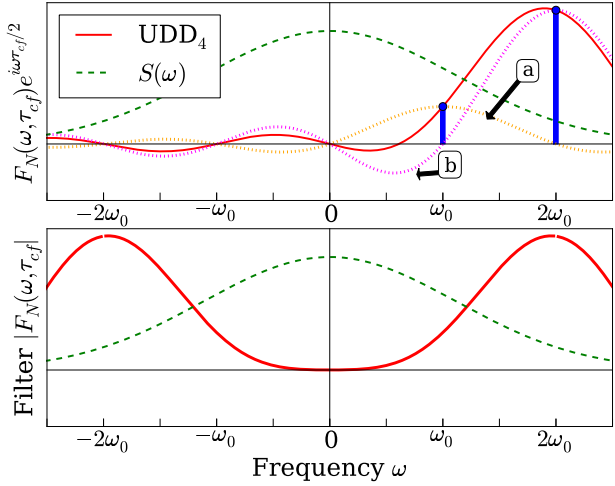}

\caption{(Color online) Decomposition of the filter function $F_{N}(\omega,\tau_{cf})$
for the UDD$_{4}$ sequence. Blue bars mark the contributions of the
harmonics of $\omega_{0}=2\pi/\tau_{cf}$. Orange and magenta dotted
lines labeled a and b are sinc functions centered at $\omega_{0}$
and $2\omega_{0}$. The top panel considers the effect only of the
first two harmonics ($k=1,2)$. They interfere to give the resulting
red thick line. The bottom panel shows how a region near to $\omega=0$
is completely canceled when we also consider the contribution of the
two harmonics with $k=-1,-2$. The green dashed line is a typical
Gaussian spectral density function of the spin-bath.}

\label{fig:pulse-harmonic2} %
\end{figure}

To determine the behavior of $F_{N}(\omega,\tau_{cf})$ close to $\omega=0$,
it is sufficient to consider the effect of only the first few harmonics
on either side of $\omega=0$. This is shown in Fig. \ref{fig:pulse-harmonic2}
for UDD$_{4}$. The red thick line in the lower panel illustrates
the stop-band for UDD$_{4}$. The blue bars are the Fourier coefficients
$A_{k}$ of the filter function $f_{4}(t^{\prime},\infty)$. The contributions
of the individual terms in Eq. (\ref{eq:FTfrequenydomain}) are represented
by the dotted lines in Fig. \ref{fig:pulse-harmonic2}. The sum ({interference})
of these diffraction effects gives rise to a maximally flat filter
shape $\left|F_{N}(\omega,\tau_{c})\right|$ close to $\omega=0$.
The top panel of Fig. \ref{fig:pulse-harmonic2} shows the effect
of the addition of diffraction effects of the first two harmonics
to the right of $\omega=0$. The resulting (red thick line) in a region
close to $\omega=0$ is canceled by the corresponding harmonics to
the left of $\omega=0$, as shown in the lower panel.

Physically therefore, the sum in Eq. (\ref{generaln}) can be understood
as the {interference} of the diffraction effects due to each inter-pulse
delay. In an optical analog, the switching between $f_{N}(t^{\prime},\tau_{cf})=1\rightarrow f_{N}(t^{\prime},\tau_{cf})=-1$
corresponds to a phase shift by $\pi$, which could be implemented
by a series of $\lambda/2$ retardation plates. For short cycle time
$\tau_{cf}\ll\tau_{B}$, the width of this region, where the filter
function vanishes, becomes broad compared to the $S(\omega)$ and
the integral (\ref{eq:decaychi}) and thus the decay tend to zero.

%

\subsection{Effect of cycle iteration: diffraction grating}

\begin{figure}
\includegraphics[scale=0.24]{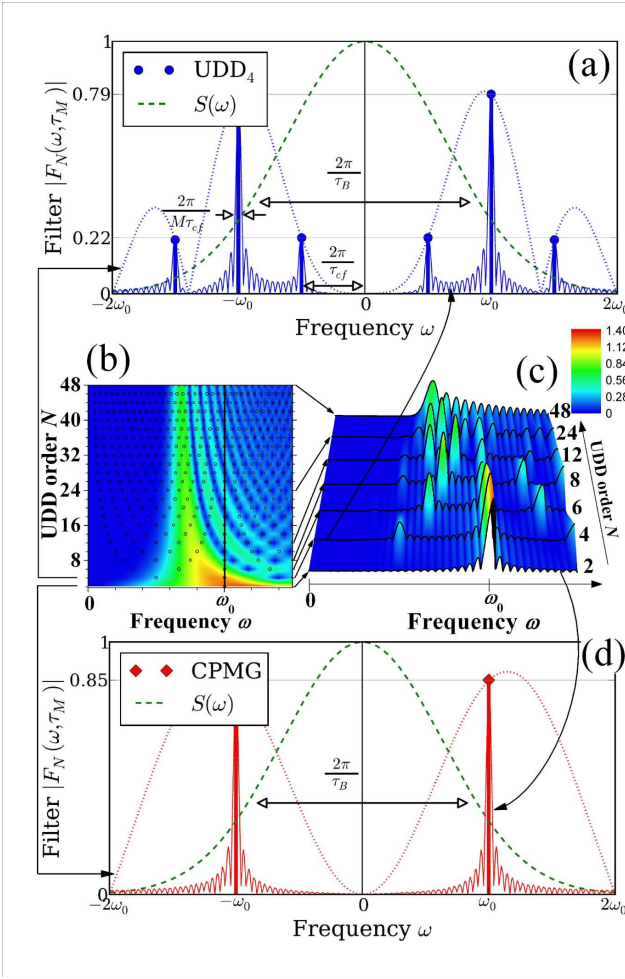} 

\caption{(Color online) Comparison of filter functions $\left|F_{N}(\omega,\tau_{M})\right|$
for different UDD orders: (a) UDD$_{4}$ with $M=1$ (dotted line)
and $M=12$ (solid line),even UDD orders $N$ with $M=1$ (b) and
$M=48/N$ (c), and (d) CPMG$=$UDD$_{2}$ with $M=1$ (dotted line)
and $M=24$ (solid line). $\omega_{0}$ is defined in terms of the
cycle time of the CPMG sequence, $\omega_{0}=2\pi/\tau_{cf}^{\mathrm{CPMG}}$.
Blue circles (a), red rhombuses (d) and empty circles in (b) are the
coefficients of the Fourier expansion of $f_{N}(t^{\prime},\infty)$.
They are modulated by the shape of the filter function $\left|F_{N}(\omega,\tau_{cf})\right|$,
shown in panel (b) and represented by blue and red dotted lines in
panels (a) and (d), where $\tau_{cf}^{CPMG}=2\tau$ and $\tau_{cf}^{\mathrm{UDD_{N}}}=N\tau$. }

\label{fig:pulse-harmonic} %
\end{figure}

Let us consider a DD sequence iterated $M$ times for a total period
$\tau_{M}=M\tau_{cf}$. This is illustrated in Fig. \ref{Flo:DDScheme}.
The corresponding filter function is then\begin{multline}
F_{N}(\omega,\tau_{M}=M\tau_{cf})=\frac{1}{\sqrt{2\pi}}\int_{0}^{\tau_{cf}}f_{N}(t^{\prime},\tau_{cf})e^{-i\omega t^{\prime}}dt^{\prime}\times\\
\times\left[1+e^{-i\omega\tau_{c}}+e^{-i2\omega\tau_{cf}}+\cdots+e^{-i(M-1)\omega\tau_{cf}}\right]\\
=\frac{\sin\left(\frac{M\omega\tau_{cf}}{2}\right)}{\sin\left(\frac{\omega\tau_{cf}}{2}\right)}e^{-\frac{i\omega(M-1)\tau_{cf}}{2}}\frac{1}{\sqrt{2\pi}}\int_{0}^{\tau_{cf}}f_{N}(t^{\prime},\tau_{cf})e^{-i\omega t^{\prime}}dt^{\prime}\\
=\frac{\sin\left(\frac{M\omega\tau_{cf}}{2}\right)}{\sin\left(\frac{\omega\tau_{cf}}{2}\right)}e^{-\frac{i\omega(M-1)\tau_{cf}}{2}}F_{N}(\omega,\tau_{cf}).\label{grating}\end{multline}
 This is analogous to the intensity pattern obtained due to diffraction
from an $M$-line grating. The maxima of the filter function $\left|F_{N}(\omega,\tau_{M})\right|$
again occur at the harmonic frequencies $\omega=2\pi k/\tau_{cf}$,
where $k$ is an integer. These are the peaks of the blue bars in
Fig. \ref{fig:pulse-harmonic2} and Fig. \ref{fig:pulse-harmonic}(a)
for UDD$_{4}$, and the red bars in Fig. \ref{fig:pulse-harmonic}(d)
for CPMG. Their amplitudes are given by the filter function $F_{N}(\omega,\tau_{cf})$
of a single cycle which is shown by blue and red dotted lines in Fig.
\ref{fig:pulse-harmonic}(a) and (b). For different even UDD orders
the single cycle filter functions $F_{N}(\omega,\tau_{cf})$ are shown
in Fig. \ref{fig:pulse-harmonic}(b) and the respective harmonic positions
are represented by the empty circles. Between two of the principal
maxima (harmonics) are $(M-2)$ secondary maxima, which are determined
by the grating transfer function $\sin(M\omega\tau_{cf}/2)/\sin(\omega\tau_{cf}/2)$.
As shown in Fig. \ref{fig:pulse-harmonic}(a) and (d) on the solid
lines and in panel (c), their amplitudes with respect to the principal
maximum falls off $\propto M^{-1}$. For example, with only 6 cycles,
the intensity of the first secondary maximum is less than 5\% of the
maximum. Note that in Fig. \ref{fig:pulse-harmonic} $\left|F_{N}(\omega,\tau_{cf})\right|$
is plotted; the contribution of the secondary maxima is drastically
reduced after taking the square.

Hence for even a few cycles, the filter function $\left|F_{N}(\omega,\tau_{M})\right|$
becomes an almost discrete spectrum that is given by the discrete
Fourier transform of $f_{N}(t^{\prime},\tau_{cf})$. This is just
the function $f_{N}(t^{\prime},\infty)$. Fig. \ref{fig:pulse-harmonic}
shows the comparison of the filter functions $F_{N}(\omega,\tau_{M})$
for UDD$_{4}$ {[}panel (a): blue solid line{]} with $M=12$ and CPMG
{[}panel (d): red solid line{]} with $M=24$. The latter choice allows
us to have the same evolution time for a fair comparison. The $\omega_{0}$
in the frequency axis is defined in terms of the cycle time for CPMG,
\begin{equation}
\omega_{0}=2\pi/(\tau_{cf}^{CPMG})=2\times2\pi/(\tau_{cf}^{\textrm{UDD}_{4}}).\end{equation}
 Figure \ref{fig:pulse-harmonic}(c) shows $F_{N}(\omega,\tau_{M})$
for different UDD orders iterated to match the same total evolution
time. In Fig. \ref{fig:pulse-harmonic2} and Fig. \ref{fig:pulse-harmonic}(a)
and (d), the spectrum corresponding to $F_{N}(\omega,\infty)$ contains
only the blue bars (for UDD$_{4}$) or the red bars (for CPMG). Equivalently
on Fig. \ref{fig:pulse-harmonic}(b), it is given by the empty circles.
The decay function $\chi(t)$ is thus a {weighted sampling} of the
spectral density function $S(\omega)$, where the weighting factor
is the magnitude of the respective Fourier components. Thus, \begin{equation}
\chi(t=\tau_{M})\propto\tau_{M}\, b_{SE}^{2}\sum_{k}A_{k}^{2}S\left(k\omega_{0}\right).\label{eq:decMult}\end{equation}
 Here, we have assumed $t=\tau_{M}\gg\tau_{B}$ so that $F(\omega,t)$
is well represented by a series of $\delta$-functions centered at
$k\omega_{0}$, given by $F_{N}(\omega,\infty)$, while the contributions
from the secondary maxima can be neglected.

This decay function (\ref{eq:decMult}) is equivalent to the one derived
by standard time dependent perturbation theory for a spin interacting
with a continuum of states. The result is similar to an expression
derived by Fermi's golden rule but the spectral density $S(\omega)$
is evaluated on the frequency components of the time-dependent perturbation\begin{equation}
\widehat{\mathcal{H}}_{SE}(t)=b_{SE}\left[\sum_{k}A_{k}\cos\left(k\omega_{0}t\right)\right]E_{z}(t)\hat{S}_{z}.\end{equation}
 It follows that for large $M$, the main contribution to the decay
rate is determined by the spectral density $S(\omega)$ at the harmonic
frequencies $k\omega_{0}$. The lowest frequency component present
in a DD sequence is the inverse of the period $\tau_{cf}$ of the
time domain filter. The CPMG sequence has the shortest period if the
average distance between pulses $\tau$ is fixed: $\tau_{cf}^{\mathrm{CPMG}}=2\tau$
and $\omega_{0}^{\mathrm{CPMG}}=\pi/(\tau)$. For an $N$ pulse UDD
cycle the period of the filter function in the time-domain is $\tau_{cf}^{\textrm{UDD}_{N}}=N\tau$
($\tau_{cf}^{\mathrm{UDD}_{N}}=2N\tau$) for $N$ even (odd). Thus
the first non-vanishing component in its Fourier expansion is at $\omega_{0}^{\mathrm{UDD}_{N}}=2\pi/(N\tau)$
($\omega_{0}^{\mathrm{UDD}_{N}}=\pi/(N\tau)$) {[}Fig. \ref{fig:pulse-harmonic}{]}.
In the system studied here, this has resulted in the CPMG generating
the slowest decay.

%

\subsection{Simulations}

For the longitudinal initial condition flip-angle errors are well
compensated as was shown in Ref. \cite{Alvarez2010}. As a consequence,
the effect of static pulse imperfections and finite pulse lengths
in the experiments does not cause a qualitative difference to the
simulations with ideal pulses. We therefore assume ideal DD pulses
and Eqs. (\ref{eq:decaychi}) and (\ref{grating}) allow us to simulate
the resulting signal decays. However, for the transverse initial condition,
the pulse errors are not compensated \cite{Alvarez2010}. Thus, here
we only describe how the simulations for the longitudinal case were
carried out.

First, the spectral density $S(\omega)$ used in Eq. (\ref{eq:decaychi})
is estimated using the FIDs of Fig. \ref{fig:fidblew}, as described
in section IV.A of Ref. \cite{Alvarez2010}. We model a finite system
to simulate the respective correlation functions that produce the
experimentally measured FIDs for both nuclei (see \cite{Alvarez2010}
for details). The {correlation time} of the bath is defined as the
time when the correlation function falls to $1/e$ of the initial
value; this was found to be around $\tau_{B}=110\mu$s.

Now, including the DD pulses, $\chi(t)$ is calculated after every
cycle period $\tau_{c}$ following Eq. (\ref{eq:decaychi}) and (\ref{grating}).
The decay rate is determined by {linearly} fitting the values of
$\chi(t)$ obtained for different times. This procedure is exact for
the case when the decay is purely exponential -- for example, when
the time $t$ far exceeds the correlation time of the bath as described
above. These simulations (lines) show reasonable agreement with the
real experimental results in Figs. \ref{fig:devmain}, \ref{fig:ordermain}
and \ref{fig:time}.

\subsection{Discussion}

\label{sec:discuss}Visualizing dynamical decoupling as a filter \cite{Cywinski2008,Biercuk2010}
for different frequency components of the environmental noise provides
a useful means for predicting relative DD performance. Some recent
discussions by Biercuk and Uys \cite{Biercuk2010} about the filter
properties of different DD sequences concerning their Fourier components
has appeared in parallel with our work. However, their work and ours
contribute from different approaches: we compare the performance of
different DD orders under the condition that the number of pulses
applied during a time interval remains constant, while Biercuk and
Uys fix the cycle time for every DD order.

Although the UDD sequence that was derived from Eq. (\ref{generaln})
has a flatter band-stop region close to $\omega=0$, it achieves this
by the interference of a {larger} number of harmonic frequencies
lower than the first harmonic of an equidistant sequence (CPMG) with
the same average spacing $\tau$ between the pulses. This is illustrated
in Fig. \ref{fig:pulse-harmonic}. The first non-zero Fourier component
of the CPMG occurs at $\omega_{0}^{\mathrm{CPMG}}=\pi/\tau$, while
for an $N$-pulse UDD sequence it occurs at $\omega_{0}^{\mathrm{UDD}_{\mathrm{even}}}=\omega_{0}^{\mathrm{CPMG}}/N$
{[}$\omega_{0}^{\mathrm{UDD}_{\mathrm{odd}}}=\omega_{0}^{\mathrm{CPMG}}/(2N)${]}.
Therefore, as one increases the UDD order, additional components appear
in the corresponding frequency-domain filter function at frequencies
below $\omega_{0}^{\mathrm{CPMG}}$.

We may thus compare the effect of two contributions to the decay rate:
The spectral density close to the lowest frequency component $\omega_{0}$
and the integral over the frequency band close to $\omega=0$. UDD
is designed to outperform CPMG in this low-frequency band and thus
superior if this is the dominant contribution. However, in the system
that we are considering here, the spectral density of the environmental
noise is still sufficiently large at $\omega_{0}^{UDD}$ that this
term dominates and leads to UDD performing worse than CPMG.

Previous works \cite{UhrigLidar2010,Hodgson2010} used perturbation
treatment of the SE interaction to predict the time evolution of the
system under UDD at short times with high precision. However, the
regimes where the decay is reduced by increasing the UDD order lead
in general to differences between their decays of order lower than
$10^{-4}$ which are too small to be determined experimentally. The
extra\-polation of those decays to longer times, by means of iterating
the DD sequences, that may result in a magnification of the UDD performance
relative to CPMG is not valid within the perturbative treatment. Our
analysis, in contrast, is valid for times $t\gg\tau_{B}$, where decay
becomes appreciable (>1\%) and thus experimentally accessible. The
filter function $F_{N}(\omega,M\tau_{cf})$ now becomes discrete when
$M$ is large and changes drastically the decay compared to the extrapolation
from a perturbative treatment of the SE interaction. From a different
approach, this situation was also observed by Khodjasteh \emph{et
al.} \cite{viola2010} for a bosonic bath: they noticed that an extrapolation
could not be done and they obtained bounds for the UDD performance
as a function of its order. Our results suggest that to achieve a
parameter range where UDD would outperform CPMG would correspond to
an almost static bath, i.e., the system-environment interaction $|b_{SE}|^{2}$
should be much stronger than the relevant intra-bath interactions.

Our model may also explain the observation in \cite{Biercuk2009}
where the odd UDD orders were found to perform substantially worse
than even orders in a qubit system consisting of $^{9}$Be$^{+}$
ions confined in a Penning trap, and where the $S(\omega)\sim1/\omega^{4}$.
This even-odd asymmetry was unexplained in their paper \cite{Biercuk2009}.

\section{Conclusions}

We have presented experimental and theoretical results evaluating
the relative performance of DD sequences with non-equidistant pulses
in a purely-dephasing spin-bath. We find that over a large range of
cycle times, number of pulses per cycle and bath-correlation time,
the equidistant sequence (CPMG) provides a measurably superior performance
over non-equidistant sequences like UDD. If the rotation axis of the
DD pulses is oriented in the direction of the initial state, we find
that the experimental results match very well with simulations using
the filter model for perfect pulses \cite{Cywinski2008}. In this
case, the effects due to the finite pulse-length, and flip-angle errors
do not qualitatively affect the results.

We interpret our results through a semi-classical model by drawing
an analogy of dynamical decoupling with interference effects in optics
and filter theory. The CPMG sequence has the shortest cycle and its
filter has the widest stop-band. By suitable interferences, the UDD
filter has the flattest stop-band close to $\omega=0$. In our system,
the position of the lowest frequency component is more important than
the very flat behavior of UDD, and thus the performance of CPMG is
better than that of non-equidistant DD sequences. We show that this
comes from the fact that the decay rate changes drastically from short
to long times compared with the bath-correlation time. While for short
evolution times, the flattest stop-band close to $\omega=0$ may play
an important role in reducing the decay rate, the fidelity reduction
is $<10^{-4}$ and thus too small to be observed experimentally. The
extrapolation of those decays to longer times that may result in a
magnification of the UDD performance relative to CPMG is not valid.
In contrast, we showed that for long times, where decay becomes appreciable,
the filter function becomes discrete, making CPMG superior because
its frequency components have the widest spacing. Conversely, we expect
that UDD should perform better in systems where the cycle time remains
significantly shorter than the bath correlation time.

Our results and the optical analogy we considered reaffirm that it
would be advantageous to tailor the DD sequence to the spectral density
function of the noise in the particular system of interest. This is
the basis behind the Locally Optimized DD (LODD) sequences \cite{Biercuk2009a,Uys2009,pan2010,pasini_optimized_2010,Biercuk2010}
or similar \cite{Gordon2008,Clausen2010}. In particular LODD can
be implemented using measurement feedback in order to find the optimal
DD sequence \cite{Biercuk2009a}. However, more work needs to be done
in methods for experimentally determining the exact noise features
in the QIP system of choice, so that techniques like LODD could be
used to determine optimal sequences for such environments. 
\begin{acknowledgments}
This work is supported by the DFG through Su 192/24-1. GAA thanks
the Alexander von Humboldt Foundation for a Research Scientist Fellowship.
We thank Götz Uhrig, Stefano Pasini and Daniel Lidar for helpful discussions,
and Marko Lovric and Ingo Niemeyer for technical support. We thank
Michael Biercuk for some useful comments about our manuscript.
\end{acknowledgments}
\bibliographystyle{apsrev} \bibliographystyle{apsrev}\bibliographystyle{apsrev}
\bibliographystyle{apsrev} \bibliographystyle{apsrev}
\bibliography{bibliography,CDD,CDD2,UDD}

\begin{thebibliography}{70}
\expandafter\ifx\csname natexlab\endcsname\relax\def\natexlab#1{#1}\fi
\expandafter\ifx\csname bibnamefont\endcsname\relax
  \def\bibnamefont#1{#1}\fi
\expandafter\ifx\csname bibfnamefont\endcsname\relax
  \def\bibfnamefont#1{#1}\fi
\expandafter\ifx\csname citenamefont\endcsname\relax
  \def\citenamefont#1{#1}\fi
\expandafter\ifx\csname url\endcsname\relax
  \def\url#1{\texttt{#1}}\fi
\expandafter\ifx\csname urlprefix\endcsname\relax\def\urlprefix{URL }\fi
\providecommand{\bibinfo}[2]{#2}
\providecommand{\eprint}[2][]{\url{#2}}

\bibitem[{\citenamefont{Nielsen and Chuang}(2000)}]{Nielsen00}
\bibinfo{author}{\bibfnamefont{M.~A.} \bibnamefont{Nielsen}} \bibnamefont{and}
  \bibinfo{author}{\bibfnamefont{I.~L.} \bibnamefont{Chuang}},
  \emph{\bibinfo{title}{Quantum Computation and Quantum Information}}
  (\bibinfo{publisher}{Cambridge University Press},
  \bibinfo{address}{Cambridge}, \bibinfo{year}{2000}).

\bibitem[{\citenamefont{Preskill}(1998)}]{Preskill1998}
\bibinfo{author}{\bibfnamefont{J.}~\bibnamefont{Preskill}},
  \bibinfo{journal}{P. Roy. Soc. Lond. A Mat.} \textbf{\bibinfo{volume}{454}},
  \bibinfo{pages}{385} (\bibinfo{year}{1998}).

\bibitem[{\citenamefont{Zurek}(2003)}]{Zurek03}
\bibinfo{author}{\bibfnamefont{W.~H.} \bibnamefont{Zurek}},
  \bibinfo{journal}{Rev. Mod. Phys.} \textbf{\bibinfo{volume}{75}},
  \bibinfo{pages}{715} (\bibinfo{year}{2003}).

\bibitem[{\citenamefont{DeChiara et~al.}(2005)\citenamefont{DeChiara, Rossini,
  Montangero, and Fazio}}]{Chiara2005}
\bibinfo{author}{\bibfnamefont{G.}~\bibnamefont{DeChiara}},
  \bibinfo{author}{\bibfnamefont{D.}~\bibnamefont{Rossini}},
  \bibinfo{author}{\bibfnamefont{S.}~\bibnamefont{Montangero}},
  \bibnamefont{and} \bibinfo{author}{\bibfnamefont{R.}~\bibnamefont{Fazio}},
  \bibinfo{journal}{Phys. Rev. A} \textbf{\bibinfo{volume}{72}},
  \bibinfo{pages}{012323} (\bibinfo{year}{2005}).

\bibitem[{\citenamefont{Burrell and Osborne}(2007)}]{Burrell2007}
\bibinfo{author}{\bibfnamefont{C.~K.} \bibnamefont{Burrell}} \bibnamefont{and}
  \bibinfo{author}{\bibfnamefont{T.~J.} \bibnamefont{Osborne}},
  \bibinfo{journal}{Phys. Rev. Lett.} \textbf{\bibinfo{volume}{99}},
  \bibinfo{pages}{167201} (\bibinfo{year}{2007}).

\bibitem[{\citenamefont{Allcock and Linden}(2009)}]{Allcock2009}
\bibinfo{author}{\bibfnamefont{J.}~\bibnamefont{Allcock}} \bibnamefont{and}
  \bibinfo{author}{\bibfnamefont{N.}~\bibnamefont{Linden}},
  \bibinfo{journal}{Phys. Rev. Lett.} \textbf{\bibinfo{volume}{102}},
  \bibinfo{pages}{110501} (\bibinfo{year}{2009}).

\bibitem[{\citenamefont{{\'{A}}lvarez and Suter}(2010)}]{AlvarezPRL2010}
\bibinfo{author}{\bibfnamefont{G.~A.} \bibnamefont{{\'{A}}lvarez}}
  \bibnamefont{and} \bibinfo{author}{\bibfnamefont{D.}~\bibnamefont{Suter}},
  \bibinfo{journal}{Phys. Rev. Lett.} \textbf{\bibinfo{volume}{104}},
  \bibinfo{pages}{230403} (\bibinfo{year}{2010}).

\bibitem[{\citenamefont{Shor}(1994)}]{Shor1994}
\bibinfo{author}{\bibfnamefont{P.~W.} \bibnamefont{Shor}}, in
  \emph{\bibinfo{booktitle}{Proceedings of the 35th Annual Symposium on the
  Foundations of Computer Science}}, edited by
  \bibinfo{editor}{\bibfnamefont{S.}~\bibnamefont{Goldwasser}}
  (\bibinfo{publisher}{IEEE Computer Society Press}, \bibinfo{address}{Los
  Alamitos, CA}, \bibinfo{year}{1994}), p. \bibinfo{pages}{124}.

\bibitem[{\citenamefont{Viola et~al.}(1999{\natexlab{a}})\citenamefont{Viola,
  Knill, and Lloyd}}]{5916}
\bibinfo{author}{\bibfnamefont{L.}~\bibnamefont{Viola}},
  \bibinfo{author}{\bibfnamefont{E.}~\bibnamefont{Knill}}, \bibnamefont{and}
  \bibinfo{author}{\bibfnamefont{S.}~\bibnamefont{Lloyd}},
  \bibinfo{journal}{Phys. Rev. Lett.} \textbf{\bibinfo{volume}{82}},
  \bibinfo{pages}{2417} (\bibinfo{year}{1999}{\natexlab{a}}).

\bibitem[{\citenamefont{Yang et~al.}(2010)\citenamefont{Yang, Wang, and
  Liu}}]{Yang2010}
\bibinfo{author}{\bibfnamefont{W.}~\bibnamefont{Yang}},
  \bibinfo{author}{\bibfnamefont{Z.}~\bibnamefont{Wang}}, \bibnamefont{and}
  \bibinfo{author}{\bibfnamefont{R.}~\bibnamefont{Liu}},
  \bibinfo{journal}{arXiv:1007.0623}  (\bibinfo{year}{2010}).

\bibitem[{\citenamefont{Taylor et~al.}(2008)\citenamefont{Taylor, Cappellaro,
  Childress, Jiang, Budker, Hemmer, Yacoby, Walsworth, and Lukin}}]{Taylor2008}
\bibinfo{author}{\bibfnamefont{J.~M.} \bibnamefont{Taylor}},
  \bibinfo{author}{\bibfnamefont{P.}~\bibnamefont{Cappellaro}},
  \bibinfo{author}{\bibfnamefont{L.}~\bibnamefont{Childress}},
  \bibinfo{author}{\bibfnamefont{L.}~\bibnamefont{Jiang}},
  \bibinfo{author}{\bibfnamefont{D.}~\bibnamefont{Budker}},
  \bibinfo{author}{\bibfnamefont{P.~R.} \bibnamefont{Hemmer}},
  \bibinfo{author}{\bibfnamefont{A.}~\bibnamefont{Yacoby}},
  \bibinfo{author}{\bibfnamefont{R.}~\bibnamefont{Walsworth}},
  \bibnamefont{and} \bibinfo{author}{\bibfnamefont{M.~D.} \bibnamefont{Lukin}},
  \bibinfo{journal}{Nat. Phys.} \textbf{\bibinfo{volume}{4}},
  \bibinfo{pages}{810} (\bibinfo{year}{2008}).

\bibitem[{\citenamefont{Hall et~al.}(2010)\citenamefont{Hall, Hill, Cole, and
  Hollenberg}}]{Hall2010}
\bibinfo{author}{\bibfnamefont{L.~T.} \bibnamefont{Hall}},
  \bibinfo{author}{\bibfnamefont{C.~D.} \bibnamefont{Hill}},
  \bibinfo{author}{\bibfnamefont{J.~H.} \bibnamefont{Cole}}, \bibnamefont{and}
  \bibinfo{author}{\bibfnamefont{L.~C.~L.} \bibnamefont{Hollenberg}},
  \bibinfo{journal}{Phys. Rev. B} \textbf{\bibinfo{volume}{82}},
  \bibinfo{pages}{045208} (\bibinfo{year}{2010}).

\bibitem[{\citenamefont{de~Lange
  et~al.}(2010{\natexlab{a}})\citenamefont{de~Lange, RistÃš, Dobrovitski, and
  Hanson}}]{Lange2010magn}
\bibinfo{author}{\bibfnamefont{G.}~\bibnamefont{de~Lange}},
  \bibinfo{author}{\bibfnamefont{D.}~\bibnamefont{RistÃš}},
  \bibinfo{author}{\bibfnamefont{V.~V.} \bibnamefont{Dobrovitski}},
  \bibnamefont{and} \bibinfo{author}{\bibfnamefont{R.}~\bibnamefont{Hanson}},
  \bibinfo{journal}{arXiv:1008.4395}  (\bibinfo{year}{2010}{\natexlab{a}}).

\bibitem[{\citenamefont{Boixo and Somma}(2008)}]{Boixo2008}
\bibinfo{author}{\bibfnamefont{S.}~\bibnamefont{Boixo}} \bibnamefont{and}
  \bibinfo{author}{\bibfnamefont{R.~D.} \bibnamefont{Somma}},
  \bibinfo{journal}{Phys. Rev. A} \textbf{\bibinfo{volume}{77}},
  \bibinfo{pages}{052320} (\bibinfo{year}{2008}).

\bibitem[{\citenamefont{Hahn}(1950)}]{Hahn1950}
\bibinfo{author}{\bibfnamefont{E.~L.} \bibnamefont{Hahn}},
  \bibinfo{journal}{Phys. Rev.} \textbf{\bibinfo{volume}{80}},
  \bibinfo{pages}{580} (\bibinfo{year}{1950}).

\bibitem[{\citenamefont{Carr and Purcell}(1954)}]{Carr1954}
\bibinfo{author}{\bibfnamefont{H.~Y.} \bibnamefont{Carr}} \bibnamefont{and}
  \bibinfo{author}{\bibfnamefont{E.~M.} \bibnamefont{Purcell}},
  \bibinfo{journal}{Phys. Rev.} \textbf{\bibinfo{volume}{94}},
  \bibinfo{pages}{630} (\bibinfo{year}{1954}).

\bibitem[{\citenamefont{Meiboom and Gill}(1958)}]{Meiboom1958}
\bibinfo{author}{\bibfnamefont{S.}~\bibnamefont{Meiboom}} \bibnamefont{and}
  \bibinfo{author}{\bibfnamefont{D.}~\bibnamefont{Gill}},
  \bibinfo{journal}{Rev. Sci. Instrum.} \textbf{\bibinfo{volume}{29}},
  \bibinfo{pages}{688} (\bibinfo{year}{1958}).

\bibitem[{\citenamefont{Viola et~al.}(1999{\natexlab{b}})\citenamefont{Viola,
  Lloyd, and Knill}}]{viola1999universal}
\bibinfo{author}{\bibfnamefont{L.}~\bibnamefont{Viola}},
  \bibinfo{author}{\bibfnamefont{S.}~\bibnamefont{Lloyd}}, \bibnamefont{and}
  \bibinfo{author}{\bibfnamefont{E.}~\bibnamefont{Knill}},
  \bibinfo{journal}{Phys. Rev. Lett.} \textbf{\bibinfo{volume}{83}},
  \bibinfo{pages}{4888} (\bibinfo{year}{1999}{\natexlab{b}}).

\bibitem[{\citenamefont{Viola and Knill}(2003)}]{viola2003}
\bibinfo{author}{\bibfnamefont{L.}~\bibnamefont{Viola}} \bibnamefont{and}
  \bibinfo{author}{\bibfnamefont{E.}~\bibnamefont{Knill}},
  \bibinfo{journal}{Phys. Rev. Lett.} \textbf{\bibinfo{volume}{90}},
  \bibinfo{pages}{037901} (\bibinfo{year}{2003}).

\bibitem[{\citenamefont{Khodjasteh and Viola}(2009)}]{Khodjasteh2009}
\bibinfo{author}{\bibfnamefont{K.}~\bibnamefont{Khodjasteh}} \bibnamefont{and}
  \bibinfo{author}{\bibfnamefont{L.}~\bibnamefont{Viola}},
  \bibinfo{journal}{Phys. Rev. Lett.} \textbf{\bibinfo{volume}{102}},
  \bibinfo{pages}{080501} (\bibinfo{year}{2009}).

\bibitem[{\citenamefont{Khodjasteh
  et~al.}(2010{\natexlab{a}})\citenamefont{Khodjasteh, Lidar, and
  Viola}}]{Khodjasteh2010}
\bibinfo{author}{\bibfnamefont{K.}~\bibnamefont{Khodjasteh}},
  \bibinfo{author}{\bibfnamefont{D.~A.} \bibnamefont{Lidar}}, \bibnamefont{and}
  \bibinfo{author}{\bibfnamefont{L.}~\bibnamefont{Viola}},
  \bibinfo{journal}{Phys. Rev. Lett.} \textbf{\bibinfo{volume}{104}},
  \bibinfo{pages}{090501} (\bibinfo{year}{2010}{\natexlab{a}}).

\bibitem[{\citenamefont{West et~al.}(2009)\citenamefont{West, Lidar, Fong,
  Gyure, Peng, and Suter}}]{west_high_2009}
\bibinfo{author}{\bibfnamefont{J.~R.} \bibnamefont{West}},
  \bibinfo{author}{\bibfnamefont{D.~A.} \bibnamefont{Lidar}},
  \bibinfo{author}{\bibfnamefont{B.~H.} \bibnamefont{Fong}},
  \bibinfo{author}{\bibfnamefont{M.~F.} \bibnamefont{Gyure}},
  \bibinfo{author}{\bibfnamefont{X.}~\bibnamefont{Peng}}, \bibnamefont{and}
  \bibinfo{author}{\bibfnamefont{D.}~\bibnamefont{Suter}},
  \bibinfo{journal}{arXiv:0911.2398}  (\bibinfo{year}{2009}).

\bibitem[{\citenamefont{West et~al.}(2010)\citenamefont{West, Fong, and
  Lidar}}]{west_near-optimal_2010}
\bibinfo{author}{\bibfnamefont{J.~R.} \bibnamefont{West}},
  \bibinfo{author}{\bibfnamefont{B.~H.} \bibnamefont{Fong}}, \bibnamefont{and}
  \bibinfo{author}{\bibfnamefont{D.~A.} \bibnamefont{Lidar}},
  \bibinfo{journal}{Phys. Rev. Lett.} \textbf{\bibinfo{volume}{104}},
  \bibinfo{pages}{130501} (\bibinfo{year}{2010}).

\bibitem[{\citenamefont{Cywinski et~al.}(2008)\citenamefont{Cywinski, Lutchyn,
  Nave, and DasSarma}}]{Cywinski2008}
\bibinfo{author}{\bibfnamefont{L.}~\bibnamefont{Cywinski}},
  \bibinfo{author}{\bibfnamefont{R.~M.} \bibnamefont{Lutchyn}},
  \bibinfo{author}{\bibfnamefont{C.~P.} \bibnamefont{Nave}}, \bibnamefont{and}
  \bibinfo{author}{\bibfnamefont{S.}~\bibnamefont{DasSarma}},
  \bibinfo{journal}{Phys. Rev. B} \textbf{\bibinfo{volume}{77}},
  \bibinfo{pages}{174509} (\bibinfo{year}{2008}).

\bibitem[{\citenamefont{Biercuk and Uys}(2010)}]{Biercuk2010}
\bibinfo{author}{\bibfnamefont{M.~J.} \bibnamefont{Biercuk}} \bibnamefont{and}
  \bibinfo{author}{\bibfnamefont{H.}~\bibnamefont{Uys}},
  \bibinfo{journal}{arXiv:1012.4262}  (\bibinfo{year}{2010}).

\bibitem[{\citenamefont{Uhrig}(2007)}]{Uhrig2007}
\bibinfo{author}{\bibfnamefont{G.~S.} \bibnamefont{Uhrig}},
  \bibinfo{journal}{Phys. Rev. Lett.} \textbf{\bibinfo{volume}{98}},
  \bibinfo{pages}{100504} (\bibinfo{year}{2007}).

\bibitem[{\citenamefont{Gordon et~al.}(2008)\citenamefont{Gordon, Kurizki, and
  Lidar}}]{Gordon2008}
\bibinfo{author}{\bibfnamefont{G.}~\bibnamefont{Gordon}},
  \bibinfo{author}{\bibfnamefont{G.}~\bibnamefont{Kurizki}}, \bibnamefont{and}
  \bibinfo{author}{\bibfnamefont{D.~A.} \bibnamefont{Lidar}},
  \bibinfo{journal}{Phys. Rev. Lett.} \textbf{\bibinfo{volume}{101}},
  \bibinfo{pages}{010403} (\bibinfo{year}{2008}).

\bibitem[{\citenamefont{Biercuk
  et~al.}(2009{\natexlab{a}})\citenamefont{Biercuk, Uys, {VanDevender}, Shiga,
  Itano, and Bollinger}}]{Biercuk2009a}
\bibinfo{author}{\bibfnamefont{M.~J.} \bibnamefont{Biercuk}},
  \bibinfo{author}{\bibfnamefont{H.}~\bibnamefont{Uys}},
  \bibinfo{author}{\bibfnamefont{A.~P.} \bibnamefont{{VanDevender}}},
  \bibinfo{author}{\bibfnamefont{N.}~\bibnamefont{Shiga}},
  \bibinfo{author}{\bibfnamefont{W.~M.} \bibnamefont{Itano}}, \bibnamefont{and}
  \bibinfo{author}{\bibfnamefont{J.~J.} \bibnamefont{Bollinger}},
  \bibinfo{journal}{Nature} \textbf{\bibinfo{volume}{458}},
  \bibinfo{pages}{996} (\bibinfo{year}{2009}{\natexlab{a}}).

\bibitem[{\citenamefont{Uys et~al.}(2009)\citenamefont{Uys, Biercuk, and
  Bollinger}}]{Uys2009}
\bibinfo{author}{\bibfnamefont{H.}~\bibnamefont{Uys}},
  \bibinfo{author}{\bibfnamefont{M.~J.} \bibnamefont{Biercuk}},
  \bibnamefont{and} \bibinfo{author}{\bibfnamefont{J.~J.}
  \bibnamefont{Bollinger}}, \bibinfo{journal}{Phys. Rev. Lett.}
  \textbf{\bibinfo{volume}{103}}, \bibinfo{pages}{040501}
  (\bibinfo{year}{2009}).

\bibitem[{\citenamefont{Clausen et~al.}(2010)\citenamefont{Clausen, Bensky, and
  Kurizki}}]{Clausen2010}
\bibinfo{author}{\bibfnamefont{J.}~\bibnamefont{Clausen}},
  \bibinfo{author}{\bibfnamefont{G.}~\bibnamefont{Bensky}}, \bibnamefont{and}
  \bibinfo{author}{\bibfnamefont{G.}~\bibnamefont{Kurizki}},
  \bibinfo{journal}{Phys. Rev. Lett.} \textbf{\bibinfo{volume}{104}},
  \bibinfo{pages}{040401} (\bibinfo{year}{2010}).

\bibitem[{\citenamefont{Pan et~al.}(2010)\citenamefont{Pan, Xi, and
  Cui}}]{pan2010}
\bibinfo{author}{\bibfnamefont{Y.}~\bibnamefont{Pan}},
  \bibinfo{author}{\bibfnamefont{Z.-R.} \bibnamefont{Xi}}, \bibnamefont{and}
  \bibinfo{author}{\bibfnamefont{W.}~\bibnamefont{Cui}},
  \bibinfo{journal}{Phys. Rev. A} \textbf{\bibinfo{volume}{81}},
  \bibinfo{pages}{022309} (\bibinfo{year}{2010}).

\bibitem[{\citenamefont{Uhrig}(2008)}]{Uhrig2008}
\bibinfo{author}{\bibfnamefont{G.~S.} \bibnamefont{Uhrig}},
  \bibinfo{journal}{New J. Phys.} \textbf{\bibinfo{volume}{10}},
  \bibinfo{pages}{083024} (\bibinfo{year}{2008}).

\bibitem[{\citenamefont{Biercuk
  et~al.}(2009{\natexlab{b}})\citenamefont{Biercuk, Uys, {VanDevender}, Shiga,
  Itano, and Bollinger}}]{Biercuk2009}
\bibinfo{author}{\bibfnamefont{M.~J.} \bibnamefont{Biercuk}},
  \bibinfo{author}{\bibfnamefont{H.}~\bibnamefont{Uys}},
  \bibinfo{author}{\bibfnamefont{A.~P.} \bibnamefont{{VanDevender}}},
  \bibinfo{author}{\bibfnamefont{N.}~\bibnamefont{Shiga}},
  \bibinfo{author}{\bibfnamefont{W.~M.} \bibnamefont{Itano}}, \bibnamefont{and}
  \bibinfo{author}{\bibfnamefont{J.~J.} \bibnamefont{Bollinger}},
  \bibinfo{journal}{Phys. Rev. A} \textbf{\bibinfo{volume}{79}},
  \bibinfo{pages}{062324} (\bibinfo{year}{2009}{\natexlab{b}}).

\bibitem[{\citenamefont{Jenista et~al.}(2009)\citenamefont{Jenista, Stokes,
  Branca, and Warren}}]{Jenista2009}
\bibinfo{author}{\bibfnamefont{E.~R.} \bibnamefont{Jenista}},
  \bibinfo{author}{\bibfnamefont{A.~M.} \bibnamefont{Stokes}},
  \bibinfo{author}{\bibfnamefont{R.~T.} \bibnamefont{Branca}},
  \bibnamefont{and} \bibinfo{author}{\bibfnamefont{W.~S.}
  \bibnamefont{Warren}}, \bibinfo{journal}{J. Chem. Phys.}
  \textbf{\bibinfo{volume}{131}}, \bibinfo{pages}{204510}
  (\bibinfo{year}{2009}).

\bibitem[{\citenamefont{Du et~al.}(2009)\citenamefont{Du, Rong, Zhao, Wang,
  Yang, and Liu}}]{Du2009}
\bibinfo{author}{\bibfnamefont{J.}~\bibnamefont{Du}},
  \bibinfo{author}{\bibfnamefont{X.}~\bibnamefont{Rong}},
  \bibinfo{author}{\bibfnamefont{N.}~\bibnamefont{Zhao}},
  \bibinfo{author}{\bibfnamefont{Y.}~\bibnamefont{Wang}},
  \bibinfo{author}{\bibfnamefont{J.}~\bibnamefont{Yang}}, \bibnamefont{and}
  \bibinfo{author}{\bibfnamefont{R.~B.} \bibnamefont{Liu}},
  \bibinfo{journal}{Nature} \textbf{\bibinfo{volume}{461}},
  \bibinfo{pages}{1265} (\bibinfo{year}{2009}).

\bibitem[{\citenamefont{Pasini and Uhrig}(2010)}]{pasini_optimized_2010}
\bibinfo{author}{\bibfnamefont{S.}~\bibnamefont{Pasini}} \bibnamefont{and}
  \bibinfo{author}{\bibfnamefont{G.~S.} \bibnamefont{Uhrig}},
  \bibinfo{journal}{Phys. Rev. A} \textbf{\bibinfo{volume}{81}},
  \bibinfo{pages}{012309} (\bibinfo{year}{2010}).

\bibitem[{\citenamefont{Abragam}(1961)}]{Abragam}
\bibinfo{author}{\bibfnamefont{A.}~\bibnamefont{Abragam}},
  \emph{\bibinfo{title}{Principles of Nuclear Magnetism}}
  (\bibinfo{publisher}{Oxford University Press, London}, \bibinfo{year}{1961}).

\bibitem[{\citenamefont{\'{A}lvarez
  et~al.}(2010{\natexlab{a}})\citenamefont{\'{A}lvarez, Ajoy, Peng, and
  Suter}}]{Alvarez2010}
\bibinfo{author}{\bibfnamefont{G.~A.} \bibnamefont{\'{A}lvarez}},
  \bibinfo{author}{\bibfnamefont{A.}~\bibnamefont{Ajoy}},
  \bibinfo{author}{\bibfnamefont{X.}~\bibnamefont{Peng}}, \bibnamefont{and}
  \bibinfo{author}{\bibfnamefont{D.}~\bibnamefont{Suter}},
  \bibinfo{journal}{Phys. Rev. A} \textbf{\bibinfo{volume}{82}},
  \bibinfo{pages}{042306} (\bibinfo{year}{2010}{\natexlab{a}}).

\bibitem[{\citenamefont{de~Lange
  et~al.}(2010{\natexlab{b}})\citenamefont{de~Lange, Wang, Riste, Dobrovitski,
  and Hanson}}]{Lange2010}
\bibinfo{author}{\bibfnamefont{G.}~\bibnamefont{de~Lange}},
  \bibinfo{author}{\bibfnamefont{Z.~H.} \bibnamefont{Wang}},
  \bibinfo{author}{\bibfnamefont{D.}~\bibnamefont{Riste}},
  \bibinfo{author}{\bibfnamefont{V.~V.} \bibnamefont{Dobrovitski}},
  \bibnamefont{and} \bibinfo{author}{\bibfnamefont{R.}~\bibnamefont{Hanson}},
  \bibinfo{journal}{Science} \textbf{\bibinfo{volume}{330}},
  \bibinfo{pages}{60} (\bibinfo{year}{2010}{\natexlab{b}}).

\bibitem[{\citenamefont{Barthel et~al.}(2010)\citenamefont{Barthel, Medford,
  Marcus, Hanson, and Gossard}}]{Barthel2010}
\bibinfo{author}{\bibfnamefont{C.}~\bibnamefont{Barthel}},
  \bibinfo{author}{\bibfnamefont{J.}~\bibnamefont{Medford}},
  \bibinfo{author}{\bibfnamefont{C.~M.} \bibnamefont{Marcus}},
  \bibinfo{author}{\bibfnamefont{M.~P.} \bibnamefont{Hanson}},
  \bibnamefont{and} \bibinfo{author}{\bibfnamefont{A.~C.}
  \bibnamefont{Gossard}}, \bibinfo{journal}{arXiv:1007.4255}
  (\bibinfo{year}{2010}).

\bibitem[{\citenamefont{Ryan et~al.}(2010)\citenamefont{Ryan, Hodges, and
  Cory}}]{Ryan2010}
\bibinfo{author}{\bibfnamefont{C.~A.} \bibnamefont{Ryan}},
  \bibinfo{author}{\bibfnamefont{J.~S.} \bibnamefont{Hodges}},
  \bibnamefont{and} \bibinfo{author}{\bibfnamefont{D.~G.} \bibnamefont{Cory}},
  \bibinfo{journal}{Phys. Rev. Lett.} \textbf{\bibinfo{volume}{105}},
  \bibinfo{pages}{200402} (\bibinfo{year}{2010}).

\bibitem[{\citenamefont{Uhrig and Lidar}(2010)}]{UhrigLidar2010}
\bibinfo{author}{\bibfnamefont{G.~S.} \bibnamefont{Uhrig}} \bibnamefont{and}
  \bibinfo{author}{\bibfnamefont{D.~A.} \bibnamefont{Lidar}},
  \bibinfo{journal}{Phys. Rev. A} \textbf{\bibinfo{volume}{82}},
  \bibinfo{pages}{012301} (\bibinfo{year}{2010}).

\bibitem[{\citenamefont{Hodgson et~al.}(2010)\citenamefont{Hodgson, Viola, and
  {D'Amico}}}]{Hodgson2010}
\bibinfo{author}{\bibfnamefont{T.~E.} \bibnamefont{Hodgson}},
  \bibinfo{author}{\bibfnamefont{L.}~\bibnamefont{Viola}}, \bibnamefont{and}
  \bibinfo{author}{\bibfnamefont{I.}~\bibnamefont{{D'Amico}}},
  \bibinfo{journal}{Phys. Rev. A} \textbf{\bibinfo{volume}{81}},
  \bibinfo{pages}{062321} (\bibinfo{year}{2010}).

\bibitem[{\citenamefont{Chen and Liu}(2010)}]{chen2010}
\bibinfo{author}{\bibfnamefont{K.}~\bibnamefont{Chen}} \bibnamefont{and}
  \bibinfo{author}{\bibfnamefont{R.-B.} \bibnamefont{Liu}},
  \bibinfo{journal}{arXiv:1009.0984}  (\bibinfo{year}{2010}).

\bibitem[{\citenamefont{Khodjasteh
  et~al.}(2010{\natexlab{b}})\citenamefont{Khodjasteh, Erdélyi, and
  Viola}}]{viola2010}
\bibinfo{author}{\bibfnamefont{K.}~\bibnamefont{Khodjasteh}},
  \bibinfo{author}{\bibfnamefont{T.}~\bibnamefont{Erdélyi}}, \bibnamefont{and}
  \bibinfo{author}{\bibfnamefont{L.}~\bibnamefont{Viola}},
  \bibinfo{journal}{arXiv:1009.1810}  (\bibinfo{year}{2010}{\natexlab{b}}).

\bibitem[{\citenamefont{Uhrig and Pasini}(2010)}]{Uhrig2010}
\bibinfo{author}{\bibfnamefont{G.~S.} \bibnamefont{Uhrig}} \bibnamefont{and}
  \bibinfo{author}{\bibfnamefont{S.}~\bibnamefont{Pasini}},
  \bibinfo{journal}{New J. Phys.} \textbf{\bibinfo{volume}{12}},
  \bibinfo{pages}{045001} (\bibinfo{year}{2010}).

\bibitem[{\citenamefont{Naydenov et~al.}(2010)\citenamefont{Naydenov, Dolde,
  Hall, Shin, Fedder, Hollenberg, Jelezko, and Wrachtrup}}]{Naydenov2010}
\bibinfo{author}{\bibfnamefont{B.}~\bibnamefont{Naydenov}},
  \bibinfo{author}{\bibfnamefont{F.}~\bibnamefont{Dolde}},
  \bibinfo{author}{\bibfnamefont{L.~T.} \bibnamefont{Hall}},
  \bibinfo{author}{\bibfnamefont{C.}~\bibnamefont{Shin}},
  \bibinfo{author}{\bibfnamefont{H.}~\bibnamefont{Fedder}},
  \bibinfo{author}{\bibfnamefont{L.~C.~L.} \bibnamefont{Hollenberg}},
  \bibinfo{author}{\bibfnamefont{F.}~\bibnamefont{Jelezko}}, \bibnamefont{and}
  \bibinfo{author}{\bibfnamefont{J.}~\bibnamefont{Wrachtrup}},
  \bibinfo{journal}{arXiv:1008.1953}  (\bibinfo{year}{2010}).

\bibitem[{\citenamefont{Bluhm et~al.}(2010)\citenamefont{Bluhm, Foletti, Neder,
  Rudner, Mahalu, Umansky, and Yacoby}}]{Bluhm2010}
\bibinfo{author}{\bibfnamefont{H.}~\bibnamefont{Bluhm}},
  \bibinfo{author}{\bibfnamefont{S.}~\bibnamefont{Foletti}},
  \bibinfo{author}{\bibfnamefont{I.}~\bibnamefont{Neder}},
  \bibinfo{author}{\bibfnamefont{M.}~\bibnamefont{Rudner}},
  \bibinfo{author}{\bibfnamefont{D.}~\bibnamefont{Mahalu}},
  \bibinfo{author}{\bibfnamefont{V.}~\bibnamefont{Umansky}}, \bibnamefont{and}
  \bibinfo{author}{\bibfnamefont{A.}~\bibnamefont{Yacoby}},
  \bibinfo{journal}{arXiv:1005.2995}  (\bibinfo{year}{2010}).

\bibitem[{\citenamefont{Hanson et~al.}(2007)\citenamefont{Hanson, Kouwenhoven,
  Petta, Tarucha, and Vandersypen}}]{Hanson2007}
\bibinfo{author}{\bibfnamefont{R.}~\bibnamefont{Hanson}},
  \bibinfo{author}{\bibfnamefont{L.~P.} \bibnamefont{Kouwenhoven}},
  \bibinfo{author}{\bibfnamefont{J.~R.} \bibnamefont{Petta}},
  \bibinfo{author}{\bibfnamefont{S.}~\bibnamefont{Tarucha}}, \bibnamefont{and}
  \bibinfo{author}{\bibfnamefont{L.~M.~K.} \bibnamefont{Vandersypen}},
  \bibinfo{journal}{Rev. Mod. Phys.} \textbf{\bibinfo{volume}{79}},
  \bibinfo{pages}{1217} (\bibinfo{year}{2007}).

\bibitem[{\citenamefont{Kane}(1998)}]{Kane1998}
\bibinfo{author}{\bibfnamefont{B.~E.} \bibnamefont{Kane}},
  \bibinfo{journal}{Nature} \textbf{\bibinfo{volume}{393}},
  \bibinfo{pages}{133} (\bibinfo{year}{1998}).

\bibitem[{\citenamefont{Morton et~al.}(2008)\citenamefont{Morton, Tyryshkin,
  Brown, Shankar, Lovett, Ardavan, Schenkel, Haller, Ager, and
  Lyon}}]{Morton2008}
\bibinfo{author}{\bibfnamefont{J.~J.~L.} \bibnamefont{Morton}},
  \bibinfo{author}{\bibfnamefont{A.~M.} \bibnamefont{Tyryshkin}},
  \bibinfo{author}{\bibfnamefont{R.~M.} \bibnamefont{Brown}},
  \bibinfo{author}{\bibfnamefont{S.}~\bibnamefont{Shankar}},
  \bibinfo{author}{\bibfnamefont{B.~W.} \bibnamefont{Lovett}},
  \bibinfo{author}{\bibfnamefont{A.}~\bibnamefont{Ardavan}},
  \bibinfo{author}{\bibfnamefont{T.}~\bibnamefont{Schenkel}},
  \bibinfo{author}{\bibfnamefont{E.~E.} \bibnamefont{Haller}},
  \bibinfo{author}{\bibfnamefont{J.~W.} \bibnamefont{Ager}}, \bibnamefont{and}
  \bibinfo{author}{\bibfnamefont{S.~A.} \bibnamefont{Lyon}},
  \bibinfo{journal}{Nature} \textbf{\bibinfo{volume}{455}},
  \bibinfo{pages}{1085} (\bibinfo{year}{2008}).

\bibitem[{\citenamefont{Dyson}(1949{\natexlab{a}})}]{dyson0}
\bibinfo{author}{\bibfnamefont{F.}~\bibnamefont{Dyson}},
  \bibinfo{journal}{Phys. Rev.} \textbf{\bibinfo{volume}{75}},
  \bibinfo{pages}{486} (\bibinfo{year}{1949}{\natexlab{a}}).

\bibitem[{\citenamefont{Dyson}(1949{\natexlab{b}})}]{dyson1}
\bibinfo{author}{\bibfnamefont{F.}~\bibnamefont{Dyson}},
  \bibinfo{journal}{Phys. Rev.} \textbf{\bibinfo{volume}{75}},
  \bibinfo{pages}{1736} (\bibinfo{year}{1949}{\natexlab{b}}).

\bibitem[{\citenamefont{Haeberlen}(1976)}]{Haeberlen1976}
\bibinfo{author}{\bibfnamefont{U.}~\bibnamefont{Haeberlen}},
  \emph{\bibinfo{title}{High Resolution NMR in Solids: Selective Averaging}}
  (\bibinfo{publisher}{Academic Press, New York}, \bibinfo{year}{1976}).

\bibitem[{\citenamefont{Magnus}(1954)}]{Magnus1954}
\bibinfo{author}{\bibfnamefont{W.}~\bibnamefont{Magnus}},
  \bibinfo{journal}{Commun. Pure Appl. Math.} \textbf{\bibinfo{volume}{7}},
  \bibinfo{pages}{649} (\bibinfo{year}{1954}).

\bibitem[{\citenamefont{Khodjasteh and
  Lidar}(2005)}]{khodjasteh_fault-tolerant_2005}
\bibinfo{author}{\bibfnamefont{K.}~\bibnamefont{Khodjasteh}} \bibnamefont{and}
  \bibinfo{author}{\bibfnamefont{D.~A.} \bibnamefont{Lidar}},
  \bibinfo{journal}{Phys. Rev. Lett.} \textbf{\bibinfo{volume}{95}},
  \bibinfo{pages}{180501} (\bibinfo{year}{2005}).

\bibitem[{\citenamefont{Haeberlen and Waugh}(1969)}]{Haeberlen1969}
\bibinfo{author}{\bibfnamefont{U.}~\bibnamefont{Haeberlen}} \bibnamefont{and}
  \bibinfo{author}{\bibfnamefont{J.~S.} \bibnamefont{Waugh}},
  \bibinfo{journal}{Phys. Rev.} \textbf{\bibinfo{volume}{185}},
  \bibinfo{pages}{420} (\bibinfo{year}{1969}).

\bibitem[{\citenamefont{Rhim et~al.}(1976)\citenamefont{Rhim, Burum, and
  Elleman}}]{Rhim1976}
\bibinfo{author}{\bibfnamefont{W.}~\bibnamefont{Rhim}},
  \bibinfo{author}{\bibfnamefont{D.~P.} \bibnamefont{Burum}}, \bibnamefont{and}
  \bibinfo{author}{\bibfnamefont{D.~D.} \bibnamefont{Elleman}},
  \bibinfo{journal}{Phys. Rev. Lett.} \textbf{\bibinfo{volume}{37}},
  \bibinfo{pages}{1764} (\bibinfo{year}{1976}).

\bibitem[{\citenamefont{Franzoni and Levstein}(2005)}]{Franzoni2005}
\bibinfo{author}{\bibfnamefont{M.~B.} \bibnamefont{Franzoni}} \bibnamefont{and}
  \bibinfo{author}{\bibfnamefont{P.~R.} \bibnamefont{Levstein}},
  \bibinfo{journal}{Phys. Rev. B} \textbf{\bibinfo{volume}{72}},
  \bibinfo{pages}{235410} (\bibinfo{year}{2005}).

\bibitem[{\citenamefont{Li et~al.}(2007)\citenamefont{Li, Dementyev, Dong,
  Ramos, and Barrett}}]{li_generating_2007}
\bibinfo{author}{\bibfnamefont{D.}~\bibnamefont{Li}},
  \bibinfo{author}{\bibfnamefont{A.~E.} \bibnamefont{Dementyev}},
  \bibinfo{author}{\bibfnamefont{Y.}~\bibnamefont{Dong}},
  \bibinfo{author}{\bibfnamefont{R.~G.} \bibnamefont{Ramos}}, \bibnamefont{and}
  \bibinfo{author}{\bibfnamefont{S.~E.} \bibnamefont{Barrett}},
  \bibinfo{journal}{Phys. Rev. Lett.} \textbf{\bibinfo{volume}{98}},
  \bibinfo{pages}{190401} (\bibinfo{year}{2007}).

\bibitem[{\citenamefont{Franzoni et~al.}(2008)\citenamefont{Franzoni, Levstein,
  Raya, and Hirschinger}}]{Franzoni2008}
\bibinfo{author}{\bibfnamefont{M.~B.} \bibnamefont{Franzoni}},
  \bibinfo{author}{\bibfnamefont{P.~R.} \bibnamefont{Levstein}},
  \bibinfo{author}{\bibfnamefont{J.}~\bibnamefont{Raya}}, \bibnamefont{and}
  \bibinfo{author}{\bibfnamefont{J.}~\bibnamefont{Hirschinger}},
  \bibinfo{journal}{Phys. Rev. B} \textbf{\bibinfo{volume}{78}},
  \bibinfo{pages}{115407} (\bibinfo{year}{2008}).

\bibitem[{\citenamefont{Li et~al.}(2008)\citenamefont{Li, Dong, Ramos, Murray,
  {MacLean}, Dementyev, and Barrett}}]{li_intrinsic_2008}
\bibinfo{author}{\bibfnamefont{D.}~\bibnamefont{Li}},
  \bibinfo{author}{\bibfnamefont{Y.}~\bibnamefont{Dong}},
  \bibinfo{author}{\bibfnamefont{R.~G.} \bibnamefont{Ramos}},
  \bibinfo{author}{\bibfnamefont{J.~D.} \bibnamefont{Murray}},
  \bibinfo{author}{\bibfnamefont{K.}~\bibnamefont{{MacLean}}},
  \bibinfo{author}{\bibfnamefont{A.~E.} \bibnamefont{Dementyev}},
  \bibnamefont{and} \bibinfo{author}{\bibfnamefont{S.~E.}
  \bibnamefont{Barrett}}, \bibinfo{journal}{Phys. Rev. B}
  \textbf{\bibinfo{volume}{77}}, \bibinfo{pages}{214306}
  (\bibinfo{year}{2008}).

\bibitem[{\citenamefont{Dong et~al.}(2008)\citenamefont{Dong, Ramos, Li, and
  Barrett}}]{dong_controlling_2008}
\bibinfo{author}{\bibfnamefont{Y.}~\bibnamefont{Dong}},
  \bibinfo{author}{\bibfnamefont{R.~G.} \bibnamefont{Ramos}},
  \bibinfo{author}{\bibfnamefont{D.}~\bibnamefont{Li}}, \bibnamefont{and}
  \bibinfo{author}{\bibfnamefont{S.~E.} \bibnamefont{Barrett}},
  \bibinfo{journal}{Phys. Rev. Lett.} \textbf{\bibinfo{volume}{100}},
  \bibinfo{pages}{247601} (\bibinfo{year}{2008}).

\bibitem[{\citenamefont{Yang and Liu}(2008)}]{Yang2008}
\bibinfo{author}{\bibfnamefont{W.}~\bibnamefont{Yang}} \bibnamefont{and}
  \bibinfo{author}{\bibfnamefont{R.}~\bibnamefont{Liu}},
  \bibinfo{journal}{Phys. Rev. Lett.} \textbf{\bibinfo{volume}{101}},
  \bibinfo{pages}{180403} (\bibinfo{year}{2008}).

\bibitem[{\citenamefont{Pines et~al.}(1972)\citenamefont{Pines, Gibby, and
  Waugh}}]{Pines72}
\bibinfo{author}{\bibfnamefont{A.}~\bibnamefont{Pines}},
  \bibinfo{author}{\bibfnamefont{M.~G.} \bibnamefont{Gibby}}, \bibnamefont{and}
  \bibinfo{author}{\bibfnamefont{J.~S.} \bibnamefont{Waugh}},
  \bibinfo{journal}{J. Chem. Phys.} \textbf{\bibinfo{volume}{56}},
  \bibinfo{pages}{1776} (\bibinfo{year}{1972}).

\bibitem[{\citenamefont{Khodjasteh and
  Lidar}(2007)}]{khodjasteh_performance_2007}
\bibinfo{author}{\bibfnamefont{K.}~\bibnamefont{Khodjasteh}} \bibnamefont{and}
  \bibinfo{author}{\bibfnamefont{D.~A.} \bibnamefont{Lidar}},
  \bibinfo{journal}{Phys. Rev. A} \textbf{\bibinfo{volume}{75}},
  \bibinfo{pages}{062310} (\bibinfo{year}{2007}).

\bibitem[{\citenamefont{Burum et~al.}(1981)\citenamefont{Burum, Linder, and
  Ernst}}]{burum1981}
\bibinfo{author}{\bibfnamefont{D.}~\bibnamefont{Burum}},
  \bibinfo{author}{\bibfnamefont{M.}~\bibnamefont{Linder}}, \bibnamefont{and}
  \bibinfo{author}{\bibfnamefont{R.}~\bibnamefont{Ernst}},
  \bibinfo{journal}{Journal of Magnetic Resonance (1969)}
  \textbf{\bibinfo{volume}{44}}, \bibinfo{pages}{173} (\bibinfo{year}{1981}).

\bibitem[{\citenamefont{Ernst et~al.}(1998)\citenamefont{Ernst, Verhoeven, and
  Meier}}]{Ernst1998}
\bibinfo{author}{\bibfnamefont{M.}~\bibnamefont{Ernst}},
  \bibinfo{author}{\bibfnamefont{A.}~\bibnamefont{Verhoeven}},
  \bibnamefont{and} \bibinfo{author}{\bibfnamefont{B.~H.} \bibnamefont{Meier}},
  \bibinfo{journal}{J. Magn. Reson.} \textbf{\bibinfo{volume}{130}},
  \bibinfo{pages}{176} (\bibinfo{year}{1998}).

\bibitem[{\citenamefont{Fiori and Pastawski}(2006)}]{Elena06}
\bibinfo{author}{\bibfnamefont{E.~R.} \bibnamefont{Fiori}} \bibnamefont{and}
  \bibinfo{author}{\bibfnamefont{H.~M.} \bibnamefont{Pastawski}},
  \bibinfo{journal}{Chem. Phys. Lett.} \textbf{\bibinfo{volume}{420}},
  \bibinfo{pages}{35} (\bibinfo{year}{2006}).

\bibitem[{\citenamefont{\'{A}lvarez
  et~al.}(2010{\natexlab{b}})\citenamefont{\'{A}lvarez, Danieli, Levstein, and
  Pastawski}}]{Alvarez2010Mesos}
\bibinfo{author}{\bibfnamefont{G.~A.} \bibnamefont{\'{A}lvarez}},
  \bibinfo{author}{\bibfnamefont{E.~P.} \bibnamefont{Danieli}},
  \bibinfo{author}{\bibfnamefont{P.~R.} \bibnamefont{Levstein}},
  \bibnamefont{and} \bibinfo{author}{\bibfnamefont{H.~M.}
  \bibnamefont{Pastawski}}, \bibinfo{journal}{Phys. Rev. A}
  \textbf{\bibinfo{volume}{82}}, \bibinfo{pages}{012310}
  (\bibinfo{year}{2010}{\natexlab{b}}).

\end{thebibliography}
 
\end{document}